\documentclass[final,journal,twoside]{IEEEtran}
\usepackage{graphicx}
\usepackage{array}
\usepackage[cmex10]{amsmath}
\usepackage{amssymb,amsthm,amsfonts}
\usepackage{array}
\usepackage{cite}
\usepackage{algorithm}
\usepackage{algorithmic}

\newtheorem{assumption}{Assumption}
\newcommand{\eg}{\emph{e.g.}}
\newcommand{\ie}{\emph{i.e.}}
\newcommand{\const}[1]{{\mathcal{#1}}}
\newcommand{\Reals}{\mathbb{R}}
\newcommand{\Complex}{\mathbb{C}}
\newcommand{\norm}[1]{\left\lVert#1\right\rVert}
\newcommand{\eqdef}{\stackrel{\Delta}{=}}
\newcommand{\der}{\mathrm{d}}
\newcommand{\zerovector}[1]{\underbrace{\begin{matrix}0 & \cdots & 0\end{matrix}}_{#1}}
\DeclareMathOperator{\diag}{diag}
\DeclareMathOperator{\sech}{sech}
\DeclareMathOperator{\sinc}{sinc}
\DeclareMathOperator{\rect}{rect}

\theoremstyle{remark}  
\newtheorem{remark}{Remark}

\begin{document}

\title{Information Transmission using the Nonlinear Fourier Transform,
  Part II:\\Numerical Methods
\thanks{\noindent Submitted for publication on April 3,
  2012; accepted December 14, 2012.  The authors are with the Edward S.\ Rogers Sr.\ Dept.\ of 
Electrical and Computer Engineering, University of Toronto, Toronto,
ON M5S 3G4, Canada, e-mails: \{mansoor,frank\}@comm.utoronto.ca.}
}

\markboth{IEEE Transactions on Information
  Theory}{Yousefi and Kschischang}

\author{Mansoor~I.~Yousefi and Frank~R. Kschischang,\IEEEmembership{~Fellow,~IEEE}}

\IEEEpubid{0000-0000/00\$00.00\copyright~2014 IEEE}
\maketitle

\begin{abstract}
\ifCLASSOPTIONonecolumn\relax\else\boldmath\fi
In this paper, numerical methods are suggested to compute the discrete
and the continuous spectrum of a signal with respect to the
Zakharov-Shabat system, a Lax operator underlying numerous integrable
communication channels including the nonlinear Schr\"odinger channel,
modeling pulse propagation in optical fibers.  These methods are
subsequently tested and their ability to estimate the spectrum are
compared against each other. These methods are used to compute the
spectrum of various signals commonly used in the optical fiber
communications. It is found that the layer-peeling and the spectral
methods are suitable schemes to estimate the nonlinear spectra with good
accuracy. To illustrate the structure of the spectrum, the locus of the
eigenvalues is determined under amplitude and phase modulation in a
number of examples. It is observed that in some cases, as signal
parameters vary, eigenvalues collide and change their course of motion.
The real axis is typically the place from which new eigenvalues
originate or are absorbed into after traveling a trajectory in the
complex plane. 
\end{abstract}
\begin{IEEEkeywords}
Optical fiber communication, forward nonlinear Fourier transform, 
Zakharov-Shabat spectral problem, numerical methods, operator eigenproblem.
\end{IEEEkeywords}

\IEEEpeerreviewmaketitle

\section{Introduction}

\IEEEPARstart{T}{he} nonlinear Fourier transform (NFT) of a signal
$q(t)$ is a pair of functions:  the \emph{continuous spectrum}
$\hat{q}(\lambda)$, $\lambda\in\Reals$, and the \emph{discrete spectrum}
$\tilde{q}(\lambda_j)$, $\Im \lambda_j>0$, $j=1,\ldots,\mathsf{N}$.  The
NFT arises in the study of integrable waveform channels as defined in
Part~I \cite{yousefi2012nft1}.  In such channels, signals propagate (in
a potentially complicated manner) according to a given integrable
evolution equation, whereas the nonlinear Fourier transform of the
signal propagates according to a (simple) multiplication operator.

In [Part~I], we proposed \emph{nonlinear frequency-division
multiplexing} (NFDM), a scheme that uses the nonlinear Fourier transform
for data communication over integrable channels.  NFDM extends
traditional orthogonal frequency-division multiplexing (OFDM) to
channels generatable by a Lax pair. An example is the optical fiber
channel, where signal propagation is modeled by the (integrable)
nonlinear Schr\"odinger (NLS) equation. In general, the channel
input-output relations in the NFT domain are (see [Part~I])
\begin{IEEEeqnarray*}{rCl}
\hat{Y}(\lambda)&=&H(\lambda)\hat{X}(\lambda) +\hat{Z},\\
\tilde{Y}(\lambda_j)&=&H(\lambda_j)\tilde{X}(\lambda_j)+\tilde{Z},
\end{IEEEeqnarray*}
where $\hat{X}(\lambda)$ and $\tilde{X}(\lambda_j)$ are continuous and
discrete spectra at the input of the channel, $\hat{Y}(\lambda)$ and
$\tilde{Y}(\lambda_j)$ are spectra at the output of the channel, and
$\hat{Z}$ and $\tilde{Z}$ represent noise.  The
channel filter $H(\lambda)$ for the NLS equation is given by
$H(\lambda)=\exp(-4j\lambda^2 z)$.

NFDM is able to deal directly with nonlinearity and dispersion, without
the need for additional compensation at the transmitter or receiver.  In
this scheme, information is encoded in the nonlinear spectrum at the
channel input, and the corresponding time-domain signal is transmitted.
At the receiver, the NFT of the received signal is computed, and the
resulting spectra $\hat{Y}(\lambda)$ and $\tilde{Y}(\lambda_j)$ are
subsequently used to recover the transmitted information.

Similar to the ordinary Fourier transform, while the NFT can be computed
analytically in a few cases, in general, numerical methods are required.
Such methods must be robust, reliable and fast enough to be implemented
in real-time at the receiver. In this paper, we suggest and evaluate the
performance of a number of numerical algorithms for computing the
forward NFT of a given signal. Using these algorithms, we then perform extensive
numerical simulations to understand the behavior of the nonlinear
spectrum for various pulse shapes and parameters commonly used in data
communications. 

\IEEEpubidadjcol
We are aware of no published work presenting the NFT of various signals
numerically, for many pulse shapes and parameters. Such work is
necessary to clarify the structure of the nonlinear spectrum and help in
its understanding. In part, this has been due to the fact that the NFT
has largely remained a theoretical artifice, and practical
implementation of the NFT as an applied tool has not yet been pursued in
engineering. 

We review the relevant literature in Section~\ref{sec:numerical-methods}
and suggest new schemes for the numerical evaluation of the forward NFT.
Although these methods are general and work for the AKNS system
\cite{ablowitz1974ist}, for the purpose of illustration, we specialize
the AKNS system to the Zakharov-Shabat system.  All these methods are
put to test in cases where analytical formulae exist and are compared with one 
another in Section~\ref{sec:test-comparison}. Only
some of these methods will be chosen for the subsequent numerical
simulations; these are the layer-peeling method, Ablowitz-Ladik
integrable discretization, and the spectral matrix eigenvalue scheme.
These methods are used in the next sections to numerically compute the
nonlinear Fourier transform of a variety of practical pulse shapes
encountered in the data communications. 

\section{The Nonlinear Fourier Transform}
\label{sec:numerical-methods}

Details of the nonlinear Fourier transform can be found in [Part~I].
Here we briefly recall a few essential ingredients required in the
numerical computation of the forward transform. As noted earlier, we
illustrate numerical methods in the context of the Zakharov-Shabat
system, which is a Lax operator for the nonlinear Schr\"odinger
equation.

For later use, we recall that the slowly-varying complex envelope
$q(t,z)$ of a narrow-band small-amplitude signal propagating in a
dispersive weakly-nonlinear medium, such as an optical fiber, satisfies
the cubic nonlinear Schr\"odinger equation. By proper scaling, the
equation can be normalized to the following dimensionless form in $1+1$
dimensions:
\begin{IEEEeqnarray}{rCl}
jq_z & = & q_{tt}+  2|q|^2 q.
\label{eq:nlsn}
\end{IEEEeqnarray}
Here $t$ denotes retarded time, and $z$ is distance.

The NFT for an integrable evolution equation starts by finding a
\emph{Lax pair} of operators $L$ and $M$ such that the evolution
equation arises as the compatibility condition $L_z=[M,L]=ML-LM$. For
the NLS equation, we may take operator $L$ as
\begin{IEEEeqnarray}{rCl}
L=j
\begin{pmatrix} 
\frac{\partial}{\partial t} & -q(t,z) \\
-q^*(t,z) & -\frac{\partial}{\partial t} 
\end{pmatrix}.
\label{eq:L-op}
\end{IEEEeqnarray}
(The corresponding $M$ operator can be found in [Part~I].)

The NFT is defined via the spectral analysis of the $L$ operator, given
in this paper by \eqref{eq:L-op}. The spectrum of $L$ is found by
solving the eigenproblem $Lv=\lambda v$, where $\lambda$ is an
eigenvalue of $L$ and $v$ is its associated eigenvector.  It can be
shown that the operator $L$ in \eqref{eq:L-op} has the \emph{isospectral
flow} property, \ie, its spectrum is invariant even as $q$
evolves according to the NLS equation.

The eigenproblem $Lv = \lambda v$ can be simplified to
\begin{IEEEeqnarray}{rCl}
v_t=
\begin{pmatrix} 
-j\lambda & q(t) \\
-q^*(t) & j\lambda 
\end{pmatrix}v.
\label{eq:dv-dt0}
\end{IEEEeqnarray}

Note that the $z$-dependence of $q$ is suppressed in \eqref{eq:dv-dt0}
(and throughout this paper), as this variable comes into play only in
the propagation of the signal, not in the definition and computation of
the NFT.

\begin{assumption}
Throughout this paper we assume that (a) $q\in L^1({\Reals})$, and
(b) $q(t)$ is supported in the finite interval $[T_1,T_2]$.  \qed
\label{ass:A}
\end{assumption}

The set of eigenvectors $v$ associated with eigenvalue $\lambda$ in
\eqref{eq:dv-dt0} is a two-dimensional subspace $E_\lambda$ of the
continuously differentiable functions.  We define the adjoint of a
vector $v=[v_1(t),v_2(t)]^T$ as $\tilde{v} = [v_2^*(t), -v_1^*(t)]^T$.
If $v(t,\lambda^*)$ is an element of $E_{\lambda^*}$, then
$\tilde{v}(t,\lambda^*)$ is an element of $E_{\lambda}$.  It can be
shown that any pair of eigenvectors $v(t,\lambda)$ and
$\tilde{v}(t,\lambda^*)$ forms a basis for $E_\lambda$ [Part~I].  Using
Assumption~\ref{ass:A}(b), we can select an eigenvector $v^1(t,\lambda)$
to be a solution of \eqref{eq:dv-dt0} with the boundary condition
\[
v^1(T_2,\lambda)=\begin{pmatrix}0\\1\end{pmatrix}e^{j\lambda T_2}.
\]
The basis eigenvectors $v^1$ and $\tilde{v}^1$ are called
\emph{canonical eigenvectors}.
 
Having identified a basis for the subspace $E_\lambda$, we can project
any other eigenvector $v^2$ on this basis according to
\begin{IEEEeqnarray}{rCl}
  \label{eq:proj}
v^2(t,\lambda)=a(\lambda)\tilde{v}^1(t,\lambda)+b(\lambda)v^1(t,\lambda).
\end{IEEEeqnarray}
Following Assumption~\ref{ass:A}(b), a particular choice for $v^2$ is
made by solving the system
\begin{IEEEeqnarray}{rCl}
v_t=
\begin{pmatrix} 
-j\lambda & q(t) \\
-q^*(t) & j\lambda 
\end{pmatrix}v,\quad
v(T_1,\lambda)=
\begin{pmatrix}1\\0\end{pmatrix}e^{-j\lambda T_1},
\IEEEeqnarraynumspace
\label{eq:dv-dt}
\end{IEEEeqnarray}
in which we dropped the superscript $2$ in $v^2$ for convenience.  By
solving \eqref{eq:dv-dt} in the interval $[T_1, T_2]$ for a given $\lambda$ and obtaining
$v(T_2,\lambda)$, the nonlinear Fourier coefficients $a(\lambda)$
and $b(\lambda)$ can be obtained by considering \eqref{eq:proj} at
$t=T_2$. The resulting coefficients obtained in this manner are
\begin{IEEEeqnarray}{rCl}
a(\lambda)&=&v_1(T_2)e^{j\lambda T_2},\IEEEyesnumber
\IEEEyessubnumber
\label{eq:ab-formula-a}
\\
b(\lambda)&=&v_2(T_2)e^{-j\lambda T_2}.\IEEEyessubnumber
\label{eq:ab-formula-b}
\end{IEEEeqnarray}

The NFT of a signal $q(t)$ consists of a continuous spectral function
defined on the real axis $\lambda\in \Reals$
\begin{IEEEeqnarray}{rCl}
\hat{q}(\lambda)=\frac{b(\lambda)}{a(\lambda)},\quad \lambda\in\Reals,
\label{eq:qhat}
\end{IEEEeqnarray}
and a discrete spectral function defined on the upper half complex plane
$\Complex^+=\{\lambda: \Im(\lambda)>0\}$
\begin{IEEEeqnarray*}{rCl}
\tilde{q}(\lambda_j)&=&\frac{b(\lambda_j)}{\der a(\lambda) /\der\lambda|_{\lambda=\lambda_j}},\: j=1,\cdots,\mathsf{N},
\end{IEEEeqnarray*}
where $\lambda_j$ are eigenvalues and correspond to the (isolated) zeros of $a(\lambda)$ in $\Complex^+$, \ie,
$a(\lambda_j)=0$. 

From the discussions made, in order to compute the nonlinear spectrum of
$q(t)$, the system of differential equations \eqref{eq:dv-dt} needs to
be solved in the interval $[T_1, T_2]$. Except for special cases,
\eqref{eq:dv-dt} needs to be solved numerically. 

Numerical methods for the calculation of the forward nonlinear Fourier transform
are divided into two classes in this article:
\begin{enumerate}
\item Methods which estimate the continuous spectrum by directly integrating
the Zakharov-Shabat system; see Section~\ref{sec:continuous}.
\item Methods which find the (discrete) eigenvalues. Two approaches are
  suggested in this paper for this purpose. Similar to the continuous spectrum
  estimation, we can integrate the Zakharov-Shabat system
  numerically and obtain $a(\lambda)$. To find zeros of $a(\lambda)$, the
  scheme is often supplemented with a search method to locate
  eigenvalues in the upper half complex plane. One can also discretize and
rewrite the Zakharov-Shabat eigenproblem $Lv=\lambda v$ in the interval $[T_1, T_2]$ as a
(large) matrix eigenvalue problem; see Section~\ref{sec:discrete}.
\end{enumerate}

We begin by discussing methods which estimate the continuous spectrum.  

\section{Numerical Methods for Computing the Continuous Spectrum}
\label{sec:continuous}

In this section, we assume that $\lambda\in\Reals$ is given and
provide algorithms for calculating the nonlinear Fourier coefficients
$a(\lambda)$ and $b(\lambda)$.  The continuous spectral function is then
easily computed from \eqref{eq:qhat}.
This process can be repeated to compute the spectral amplitudes for any
desired finite set of continuous frequencies $\lambda$.

\subsection{Forward and Central Discretizations}

The most obvious method to attempt to solve \eqref{eq:dv-dt} is the
first-order Euler method or one of its variations \cite{stoer1993ina}.

Recall that the signal $q(t)$ is supported in the finite time interval
$[T_1,T_2]$, and partition this interval uniformly according to the mesh
$T_1<T_1+\epsilon <\cdots<T_1+N\epsilon=T_2$ with size $N$, \ie,
with $\epsilon = (T_2-T_1)/N$.
Let $q[k]\eqdef q(T_1+k\epsilon)$
and let
\begin{IEEEeqnarray}{rCl}
P[k] \eqdef \begin{pmatrix} -j\lambda & q[k] \\-q^*[k] & j\lambda \end{pmatrix}.
\label{eq:pk}
\end{IEEEeqnarray}
Integrating both sides of \eqref{eq:dv-dt} from
$k\epsilon$ to $(k+1)\epsilon$ and assuming that the right hand side is
constant over this interval, we get
\begin{IEEEeqnarray}{rCl}
v[k+1]& = &A_f[k]v[k], \quad k=0,\ldots,N,\nonumber \\
v[0]  & = & \begin{pmatrix} 1 \\ 0 \end{pmatrix}e^{-j\lambda T_1},
\label{eq:euler}
\end{IEEEeqnarray}
where $A_f[k]=I_2+\epsilon P[k]$ and $I_2$ is the identity matrix. Equation \eqref{eq:euler} 
is iterated from $k=0$ to $k=N$ to find $v[N]$.  The resulting vector is subsequently substituted in
\eqref{eq:ab-formula-a}--\eqref{eq:ab-formula-b} to obtain $a(\lambda)$ and $b(\lambda)$. 

We have implemented the Euler method for the calculation of the
nonlinear Fourier transform of a number of pulse shapes. Unfortunately,
the one-step Euler method does not produce satisfactory results for
affordable small step sizes $\epsilon$.

One can improve upon the basic Euler method by  considering the
\emph{central-difference iteration} \cite{stoer1993ina},
\begin{IEEEeqnarray*}{rCl}
v[k+1]&=&v[k-1]+2\epsilon P[k]v[k].
\end{IEEEeqnarray*}
This makes the discretization second-order, \ie, the error
$v(T_1+k\epsilon)-v[k]$ is of order $\mathcal{O}(\epsilon^2)$.  Here an
additional initial condition is required too, which can be obtained,
\eg, by performing one step of the regular forward difference
\eqref{eq:euler}. 

\subsection{Fourth-Order Runge-Kutta Method}

One can also employ higher-order integration schemes such as the
Runge-Kutta methods. Improved results are obtained using the
fourth-order Runge-Kutta method \cite{mclaughlin1992chb,
olivier2008nss}. However it takes significant time to estimate the
spectrum using such higher-order numerical methods in real-time. Since
the method, with its typical parameters, is quite slow and does
not outperform some of the schemes suggested in the following sections, we do
not elaborate on this method here; see \cite{stoer1993ina} for
details. However, for comparison purposes we will include this scheme in our numerical
simulations given in Section~\ref{sec:test-comparison}.

\subsection{Layer-Peeling Method}
\label{sec:layer-peeling}
In Section IV.~C of [Part~I], we have calculated the nonlinear spectra
of a rectangular pulse.  One can approximate $q(t)$ as a piece-wise
constant signal and use the layer-peeling property of the nonlinear
Fourier transform to estimate the spectrum of any given signal. Let
$a[k]$ and $b[k]$ be the nonlinear Fourier coefficients of $q(t)$ in
the interval $[T_1, T_1+k\epsilon)$, and $x[k]$ and $y[k]$ coefficients in
the small (rectangular) region
$[T_1+k\epsilon,T_1+k\epsilon+\epsilon)$. The iterations of the layer-peeling
method read
\begin{IEEEeqnarray}{rCl}
(a[k+1],b[k+1])& = & (a[k],b[k])\circ(x[k],y[k]),  
\label{eq:layer-peeling} \\
(a[0],b[0]) & = & (1,0), \nonumber 
\end{IEEEeqnarray}
where the $\circ$ operation is defined as in \cite{yousefi2012nft1}
\begin{IEEEeqnarray*}{rCl}
a[k+1]=a[k]x[k]-b[k]\bar{y}[k],\nonumber\\
b[k+1]=a[k]y[k]+b[k]\bar{x}[k],
\end{IEEEeqnarray*}
in which
\begin{IEEEeqnarray}{rCl}
x[k]&=&\left(\cos(D\epsilon )-j\frac{\lambda}{D}\sin(D\epsilon )\right)e^{j\lambda \left(t[k]-t[k-1]\right)}, \nonumber \\
y[k]&=& \frac{-q^*[k]}{D}\sin(D\epsilon)e^{-j\lambda \left(t[k]+t[k-1]\right)},
\label{eq:xk-yk}
\end{IEEEeqnarray}
and $\bar{x}[k](\lambda)=x^*[k](\lambda^*)$,
$\bar{y}[k](\lambda)=y^*[k](\lambda^*)$,
$D=\sqrt{\lambda^2+|q[k]|^2}$. The desired coefficients are obtained
as $a:=a[N]$ and $b:=b[N]$. Note that the exponential factors
in $x[k]$ and $y[k]$ enter \eqref{eq:xk-yk} in a telescopic manner. As a result, for the numerical
implementation, it is faster to drop these factors and just scale the resulting $a[N]$ and $b[N]$ coefficients 
by $\exp(j\lambda(T_2-T_1))$ and $\exp(-j\lambda(T_2+T_1))$,
respectively. This, however, reduces the accuracy as it involves the
product of large and small numbers.
  
We are motivated by \cite{tao2006nfa} in which the layer-peeling
identity \eqref{eq:layer-peeling} is mentioned as a property of the
nonlinear Fourier transform. An equivalent presentation of this method
is given in \cite{osborne2009now} as well.

Note further that a different numerical method, but with the same name
(layer-peeling), exists in geophysics and fiber Bragg design
\cite{skaar2003design};  however this method is not directly related to the
\emph{forward} NFT problem considered here.

We shall see in Section~\ref{sec:test-comparison} that the layer-peeling
method gives remarkably accurate results in estimating the nonlinear
Fourier transform.  

\subsection{Crank-Nicolson Method}

In the Crank-Nicolson method, the derivative of the evolution
parameter is approximated by a finite-difference approximation, \eg,
forward discretization, and other functions are discretized by taking
their average over the end points of the discretization interval:
\begin{IEEEeqnarray*}{rCl}
\frac{v[k+1]-v[k]}{\epsilon}=\frac{1}{2}\left(P[k] v[k]+P[k+1]
  v[k+1]\right), 
\end{IEEEeqnarray*}
where $P[k]$ is defined in \eqref{eq:pk}. This implicit iteration
can
be made explicit 
\begin{IEEEeqnarray*}{rCl}
v[k+1]=A_{cn}v[k], \quad k=0,\cdots, N,
\end{IEEEeqnarray*}
with the initial condition \eqref{eq:euler}, where
\begin{IEEEeqnarray*}{rCl}
A_{cn}=(I-\frac{\epsilon}{2}P[k+1])^{-1}(I+\frac{\epsilon}{2}P[k]).
\end{IEEEeqnarray*}
As we will see, this simple
scheme too gives good results in estimating the nonlinear spectrum.

\subsection{The Ablowitz-Ladik Discretization}

Ablowitz-Ladik (AL) discretization is an integrable discretization of the NLS
equation in time domain \cite{ablowitz1976ndd}. In this section, we suggest using the
Lax pairs of the Ablowitz-Ladik discretization of the NLS equation for
solving the Zakharov-Shabat eigenproblem in the spectral domain. 

Discretization sometimes breaks symmetries, making the discrete version
of an integrable equation no longer integrable. A consequence of
symmetry-breaking is that quantities which are conserved in the
continuous model may no longer be invariant in the discrete model.
A completely integrable Hamiltonian system with an infinite number of
conserved quantities might have a discretized version with no, or few,
conserved quantities. The discrete equation therefore does not quite
mimic the essential features of the original equation if the step size
is not small enough. 

However, for some integrable equations, discretizations exist which are
themselves completely integrable Hamiltonian systems, \ie, they possess an
infinite number of conserved quantities and are linearizable by a Lax
pair, and therefore are solvable by the nonlinear Fourier transform. Such
developments exist for the NLS and Korteweg-de~Vries (KdV) equations.

For the NLS equation, the integrable discrete version was introduced by
Ablowitz and Ladik \cite{ablowitz1976ndd}. To illustrate the general
idea, let us replace $1\pm j\lambda\epsilon$ for small $\epsilon$ with
$e^{\pm j\lambda\epsilon}$ in the forward discretization method \eqref{eq:euler} (the
opposite of what is usually done in practice).  Let
$z=e^{-j\lambda\epsilon}$ represent the discrete eigenvalue,
$Q[k]=q[k]\epsilon$, and 
\begin{IEEEeqnarray}{rCl}
A_{al}[k]=
\begin{pmatrix}
 z & Q[k] \\
-Q^*[k] & z^{-1}
\end{pmatrix}.
\label{eq:AL-R}
\end{IEEEeqnarray}
The Ablowitz-Ladik iteration is
\begin{IEEEeqnarray}{rCl}
v[k+1]=A_{al}[k]v[k],\quad  v[0]=
\begin{pmatrix}
1\\
0
\end{pmatrix}e^{-j\lambda T_1}.
\label{eq:al-iter}
\end{IEEEeqnarray}
Under the $z$ transformation, the upper half complex plane in $\lambda$
domain is mapped to the exterior of the unit circle in the $z$ domain.
The continuous spectrum therefore lies on the unit circle $|z|=1$, while the
discrete spectrum lies outside of the unit circle $|z|>1$.   

One can rewrite the $R$-equation \eqref{eq:AL-R} in the eigenvalue
form $Lv[k]=zv[k]$, with the following $L$ operator 
\begin{IEEEeqnarray}{rCl}
L=
\begin{pmatrix}
\mathcal{Z} & -Q[k] \\
-Q^*[k-1] & \alpha[k-1]\mathcal{Z}^{-1}
\end{pmatrix},
\label{eq:AL-L1}
\end{IEEEeqnarray}
where $\alpha[k]=1+|Q[k]|^2$, and $\mathcal{Z}$ is the shift operator, \ie,
$\mathcal{Z}^{-1}x[k]=x[k-1]$, $k\in\mathbb{Z}$. To the first order in
$\epsilon$, $\alpha[k]\approx 1$ and \eqref{eq:AL-L1} can be simplified to
\begin{IEEEeqnarray}{rCl}
L=
\begin{pmatrix}
\mathcal{Z} & -Q[k] \\
-Q^*[k] & \mathcal{Z}^{-1}
\end{pmatrix}.
\label{eq:AL-L}
\end{IEEEeqnarray}
Given the $L$ operator \eqref{eq:AL-L}, one can consider the $M$ operator
of the continuous NLS equation and modify its elements such that the compatibility 
equation $L_z=[M,L]$ represents a
discretized version of the NLS equation. It is not hard to verify that
after doing so we are led to an $M$ operator resulting in the
following discrete integrable NLS equation \cite{ablowitz1976ndd, ablowitz2006sai}
\begin{IEEEeqnarray}{rCl}
j\frac{\der q[k]}{\der z}&=&\frac{q[k+1]-2q[k]+q[k-1]}{\epsilon^2}\nonumber \\
&&+\:|q[k]|^2(q[k+1]+q[k-1]).
\label{eq:disc-nls}
\end{IEEEeqnarray}
Here the space derivative remains intact and the signal $q[k]$ is
discretized in time, in such a way that the nonlinearity is somewhat
averaged among three time samples. In the continuum limit
$\epsilon\rightarrow 0$, \eqref{eq:disc-nls} approaches the continuous
NLS equation and its merits lie in the fact that it is integrable for
any $\epsilon$, not just in the limit $\epsilon
\rightarrow 0$. For example, soliton signals can be
observed in this model for any $\epsilon$. The equation has its own infinite number
constants of motion, approaching integrals of motion in the continuum
limit. The operator $M$ which leads to \eqref{eq:disc-nls}, and the
details of the nonlinear Fourier transform for \eqref{eq:disc-nls} can be found in
\cite{ablowitz1976ndd, ablowitz2006sai}. 

We conclude that the Ablowitz-Ladik discretization can be used not only
as a means to discretize the NLS equation in the time domain \cite{taha1984ana}, but also as a
means to solve the continuous-time Zakharov-Shabat system in the
spectral domain. This is a non-finite-difference discretization, capable
of dealing with oscillations $\exp(\pm j\lambda t)$ in the
Zakharov-Shabat system, which greatly enhances the accuracy of the
one-step finite-difference methods.

Following Tao and Thiele's approach \cite{tao2006nfa} and
\cite{mclaughlin1992chb}, we can also normalize the $A_{al}[k]$ matrix  
\begin{IEEEeqnarray}{rCl}
v[k+1]=\frac{1}{\sqrt{1+|Q[k]|^2}}
\begin{pmatrix}
z & Q[k] \\
-Q[k]^* & z^{-1}
\end{pmatrix}v[k].
\label{eq:al-iter2}
\end{IEEEeqnarray}
The scale factor does not change the spectrum significantly, since it is canceled out
in the ratios $\hat{q}=b/a$ and $\tilde{q}=b/a^\prime$, and also its
effects are second-order in $\epsilon$. However, numerically, normalization may
help in reducing the numerical error. 
In subsequent sections, we refer to \eqref{eq:al-iter}
as the Ablowitz-Ladik method (AL1) and to \eqref{eq:al-iter2} as the 
modified Ablowitz-Ladik method (AL2).

\section{Methods for Calculating the Discrete Spectrum}
\label{sec:discrete}

In order to compute the discrete spectrum, the zeros of $a(\lambda)$ in
the upper half complex plane must be found.  One way to visualize this
is to assume a two-dimensional mesh in $\mathbb{C}^+$ and determine
$a$ at all mesh points.  Discrete eigenvalues are then easily
identified by looking at the graph of $|a(\lambda)|$; in many cases they correspond to
deep and narrow ``wells'' corresponding to the zeros of the magnitude of
$a$. 

As noted earlier, two types of methods are suggested to calculate the point
spectrum.
\begin{enumerate}
\item One can use the integration-based algorithms mentioned in Section~\ref{sec:continuous} 
which calculate nonlinear Fourier coefficients, and search 
for eigenvalues using a root 
finding method, such as the Newton-Raphson method. Such methods require
good initial points and one needs to be careful about convergence
\cite{stoer1993ina}.
\item It is also possible to rewrite the spectral problem for an operator as a (large) matrix 
eigenvalue problem. The point spectrum of the operator can be found in this way too. 
\end{enumerate}

\subsection{Discrete Spectrum via Search Methods}
\label{sec:search-methods}
To calculate the discrete spectral amplitude $\tilde{q}=b/a^\prime$,
we require
$\der a/\der \lambda$ as well.
As we will show,
information about
the  derivative of $a$ can be updated recursively along with the
information about $a$, without resorting to approximate numerical differentiation.

Recall that the nonlinear Fourier coefficient $a(\lambda)$ is given by \eqref{eq:ab-formula-a} 
\begin{IEEEeqnarray*}{rCl}
a(\lambda)&=&v_1[N]e^{j\lambda t[N]}.
\end{IEEEeqnarray*}
Taking the derivative with respect to $\lambda$, we obtain
\begin{IEEEeqnarray*}{rCl}
\frac{\der a(\lambda)}{\der\lambda}&=&\left((v_1[N])^\prime+j
  t[N]v_1[N]\right) e^{j\lambda t[N]}.
\end{IEEEeqnarray*}
We can update the derivative information $\der v/\der \lambda$ along with $v$.
In methods of Section~\ref{sec:continuous}, the transformation of eigenvectors from $t[k]$ to $t[k+1]$ can
be generally represented as 
\begin{IEEEeqnarray*}{rCl}
v[k+1]=A[k]v[k],
\end{IEEEeqnarray*} 
for some suitable one-step update matrix $A[k]$, which varies from method to
method. Differentiating with respect to 
$\lambda$ and augmenting $v$ with $v^\prime=\der v/\der\lambda$, we get the iterations
 
\begin{IEEEeqnarray}{rCl}
v[k+1]&=&A[k]v[k],\IEEEyesnumber\IEEEyessubnumber\label{eq:v-vprime-1}\\
v^\prime[k+1]&=&A^\prime[k]v[k]+A[k]v^\prime[k],\IEEEyessubnumber
\label{eq:v-vprime-2}
\end{IEEEeqnarray} 
with initial conditions
\begin{IEEEeqnarray*}{rCl}
v[0]
=\begin{pmatrix}
1\\
0\\
\end{pmatrix}
e^{-j\lambda t[0]},
\quad
v^\prime[0]=
\begin{pmatrix}
-jt[0]\\
0
\end{pmatrix}
e^{-j\lambda t[0]}.
\end{IEEEeqnarray*}
The derivative matrix $A^\prime$ depends on the method used.

For the forward discretization scheme:
\begin{IEEEeqnarray*}{rCl}
A^\prime_{f}=M_1=
\begin{pmatrix}
-j & 0\\
0 & j
\end{pmatrix}\epsilon.
\end{IEEEeqnarray*}

For the Crank-Nicolson scheme: 
\[
A^\prime_{cn}=\frac{\epsilon}{2}\left(I-\frac{\epsilon}{2}P[k+1]\right)^{-1}\Sigma_3 \left(I+A_{cn}[k]\right),
\]
where $\Sigma_3=\diag(-j, j)$.

For the Ablowitz-Ladik method:
\begin{IEEEeqnarray*}{rCl}
A^\prime_{al}=
\begin{pmatrix}
-jz & 0\\
0 & jz^{-1}
\end{pmatrix}\epsilon.
\end{IEEEeqnarray*}

The desired coefficients are obtained at $k=N$ as follows
\begin{IEEEeqnarray*}{rCl}
a(\lambda)&=&v_1[N]e^{j\lambda t[N]},\\
b(\lambda)&=&v_2[N]e^{-j\lambda t[N]},\\
a^\prime(\lambda)&=&\left(v^\prime_1[N]+jt[N]v_1[N]\right)e^{j\lambda t[N]}.
\end{IEEEeqnarray*}

Similarly, the layer-peeling iteration can be augmented to
update $a^\prime(\lambda)$
as well:
\begin{IEEEeqnarray*}{rCl}
a^\prime[k+1]&=&a^\prime[k]x[k]+a[k]x^\prime [k]-(b^\prime[k]\bar{y}[k]+b[k]\bar{y}^\prime[k]),\nonumber\\
b^\prime[k+1]&=&a^\prime[k]y[k]+a[k]y^\prime [k]+b^\prime[k]\bar{x}[k]+b[k]\bar{x}^\prime [k],\nonumber\\
a^\prime[0]&=&b^\prime[0]=0,
\end{IEEEeqnarray*}
where 
\begin{IEEEeqnarray*}{rCl}
x^\prime[k]&=&j\epsilon\left(1-\frac{\lambda^2}{D^2}\right)\left(\cos(D\epsilon)-\frac{\sin(D\epsilon)}{D\epsilon}\right)e^{j\lambda\epsilon},\\
y^\prime[k]&=& -q^*[k]\bigg\{\frac{\lambda\epsilon}{D^2}\cos(D\epsilon)\\
&&-\:\Bigl(\frac{\lambda}{D^3}+j\frac{t[k]+t[k-1]}{D}\Bigr)\sin(D\epsilon) \bigg\}e^{-j\lambda(t[k]+t[k-1])}.
\end{IEEEeqnarray*}
The expressions for $\bar{x}^\prime[k]$ and $\bar{y}^\prime[k]$ are
similar, with all $j$'s replaced with $-j$
and $q^*[k]$ replaced with $q[k]$.
 
With the derivative information being available, the Newton-Raphson method is a
good scheme to search for the location of the (discrete) eigenvalues. The
iteration for the complex-valued Newton-Raphson scheme is
\begin{IEEEeqnarray}{rCl}
\lambda_{k+1}=\lambda_k-\alpha_k\frac{a(\lambda_k)}{a^\prime(\lambda_k)},
\label{eq:newton}
\end{IEEEeqnarray}
where $\alpha_k$ is some step size modifier; usually $\alpha_k=1$. The
iteration stops if $\lambda_k$ is almost stationary, \ie, if
$|\alpha\frac{a}{a^\prime}|<\delta$ for a small $\delta$. In practice,
the quadratic convergence of the scheme is often very fast and occurs in
just a few iterations.

In data communications, since noise is usually small, the points in the
transmitted constellation can (repeatedly) serve as the initial
conditions for \eqref{eq:newton}. In this case, convergence is usually
achieved in a couple of iterations. For an unknown signal, random initial
conditions are chosen. In either case, one or more sequence of Newton
iterations 
have to be performed for any single eigenvalue. 

To make sure that all of the eigenvalues are found,
we can check the trace formula for $n=1,2,3$ [Part I].  The trace formula is a time frequency identity
relating the hierarchy of infinitely many conserved quantities to
the spectral components. 

In general, the trace formula represents a time-domain conserved quantity
as the sum of discrete and continuous spectral terms:
\begin{IEEEeqnarray*}{rCl}
E^{(k)}=E_{\textnormal{disc}}^{(k)}+E_{\textnormal{cont}}^{(k)},\quad k=0,1, \cdots ,
\end{IEEEeqnarray*}
where
\begin{IEEEeqnarray*}{rCl}
E_{\textnormal{disc}}^{(k)}&=&\frac{4}{k}\sum\limits_{i=1}^{\mathsf{N}}\Im\left(\lambda_i^k\right),
\\
E_{\textnormal{cont}}^{(k)}&=&
\frac{1}{\pi}\int\limits_{-\infty}^{\infty}\lambda^{k-1}\log\left(1+|\hat{q}(\lambda)|^2\right)d\lambda,
\end{IEEEeqnarray*}
and where $E^{(k)}$ are time-domain conserved quantities (functionals of
the signal). The first few conserved quantities are energy
\begin{IEEEeqnarray*}{rCl}
E^{(1)}=\int\limits_{-\infty}^{\infty}|q(t,z)|^2dt,
\end{IEEEeqnarray*}
momentum
\begin{IEEEeqnarray*}{rCl}
E^{(2)}=\frac{1}{2j} \int\limits_{-\infty}^{\infty}q(t)\frac{\der q^*(t)}{\der t}dt,
\end{IEEEeqnarray*}
and Hamiltonian
\begin{IEEEeqnarray*}{rCl}
E^{(3)}=\frac{1}{(2j)^2} \int\limits_{-\infty}^{\infty}\left(\left|q(t)\right|^4-\left|\frac{\der q(t)}{\der t}\right|^2\right)\der t.
\end{IEEEeqnarray*}
For $n=1$, the trace formula is a kind of Parseval's identity
(Plancherel's theorem), representing the total energy of the signal in time as the sum of the
energy of the discrete and continuous spectral functions. When satisfied, the Parseval's identity ensures that all of the signal energy has been accounted for.

We first calculate the continuous spectrum and its ``energy terms'' for a
sufficiently fine mesh on the real $\lambda$ axis. The norm of the error vector
\begin{IEEEeqnarray}{rCl}
E_{\textnormal{error}}=\left(
E^{(1)}-E_{\textnormal{cont}}^{(1)},\:
E^{(2)}-E_{\textnormal{cont}}^{(2)},\:
E^{(3)}-E_{\textnormal{cont}}^{(3)}\right),
\IEEEeqnarraynumspace
\label{eq:error-vec}
\end{IEEEeqnarray}
gives an estimate on
the number of remaining eigenvalues. When a new eigenvalue $\lambda$ is found, this
error is updated as 
\[
E_{\textnormal{error}}:=E_{\textnormal{error}}- \left(4\Im\lambda,\:
2\Im\lambda^2,\:
\frac{4}{3}\Im\lambda^3
\right).
\]
The process is repeated until
$\norm{E_{\textnormal{error}}}$ is less than a small prescribed tolerance
value. 

To summarize, given the signal $q(t)$, its nonlinear Fourier transform
can be computed based on Algorithm 1.

\begin{algorithm}
\caption{Numerical Nonlinear Spectrum Estimation}

\begin{algorithmic}
\STATE Sample the signal at a sufficiently small step size $\epsilon$.  
\STATE Fix a sufficiently fine mesh $M$ on the real $\lambda$ axis.
\FOR{ each $\lambda\in M$}
\STATE Iterate \eqref{eq:v-vprime-1} from $k=0$ to $k=N$ to obtain $v[N]$.
\STATE Compute the continuous spectral amplitude 
\begin{IEEEeqnarray*}{rCl}
\hat{q}(\lambda)=\frac{v_2[N]}{v_1[N]}e^{-2j\lambda t[N]}.
\end{IEEEeqnarray*} 
\ENDFOR
\STATE Initialize the error $e=\norm{E_{\textnormal{error}}}$, according to \eqref{eq:error-vec}.
\WHILE{$e>\epsilon_1$}
\STATE Choose $\lambda_0\in\mathcal{D}$ randomly, where $\mathcal{D}$
is a prescribed region in $\Complex^+$. 
\STATE Set $i=0$; 
\REPEAT
\STATE Iterate \eqref{eq:v-vprime-1}--\eqref{eq:v-vprime-2} from $k=0$ to $k=N$ to obtain $v[N]$ and $v^\prime[N]$, and 
perform a Newton-Raphson update
\begin{IEEEeqnarray*}{rCl}
\lambda_{i+1}=\lambda_{i}-\Delta\lambda,\quad\Delta\lambda=\alpha_i \frac{v_1[N]}{v_1^\prime[N]+jt[N]v_1[N]}.
\end{IEEEeqnarray*}
\IF{ $\lambda_{i+1}\notin \mathcal{D}$}
\STATE Choose $\lambda_0\in\mathcal{D}$ randomly.
\STATE Set $i:=-1$.
\ENDIF
\STATE Set $i:=i+1$.
\UNTIL{$|\Delta\lambda|<\epsilon_2$ and $i>0$}
\STATE $\lambda_i$ is an eigenvalue and the associated spectral amplitude is 
\begin{IEEEeqnarray*}{rCl}
\tilde{q}(\lambda_i)=\frac{v_2[N]}{v_1^\prime[N]+jt[N]v_1[N]}e^{-2j\lambda_{i} t[N]}.
\end{IEEEeqnarray*} 
\STATE Update the error
$E_{\textnormal{error}}:= E_{\textnormal{error}}-E_{\textnormal{disc}}^i$, $e=\norm{E_{\textnormal{error}}}$, where $E_{\textnormal{disc}}^i=\left(4\Im\lambda_i, 2\Im\lambda^2_i, \frac{4}{3}\Im\lambda^3_i\right)$. 
\ENDWHILE
\end{algorithmic}
\end{algorithm}

\subsection{Discrete Spectrum as a Matrix Eigenvalue Problem}

The methods mentioned in Section~\ref{sec:search-methods} find the
discrete spectrum by searching for eigenvalues in the upper half complex plane.
Sometimes it is desirable to have all eigenvalues at once, which can be
done by solving a matrix eigenvalue problem \cite{olivier2008nss}. These
schemes obviously estimate only (discrete) eigenvalues and do not give
information on the rest of the spectrum.  Since the matrix eigenvalue
problem can be solved quickly for small-sized problems, it might take
less computational effort to compute the discrete spectrum in this way. In
addition, one does not already need the continuous spectrum to estimate
the size of the discrete spectrum.  On the other hand, for large
matrices that arise when a large number of signal samples are used, the
matrix eigenproblem (which is not Hermitian) is slow and it may
be better to find the discrete spectrum using the search-based methods.
The matrix-based methods also have the disadvantages that they can
generate a large number of spurious eigenvalues, and one may not be able to
restrict the algorithm for finding eigenvalues of a matrix to a certain region of the
complex plane.

Below we rewrite some of the methods mentioned in the
Section~\ref{sec:continuous} as a regular matrix eigenvalue problem,
for the computation of the discrete spectrum.

\subsubsection{Central-Difference Eigenproblem}
The matrix eigenvalue problem can be formulated in the time domain or
the  frequency
domain \cite{olivier2008nss}. Consider the Zakharov-Shabat system in
the form
$Lv=\lambda v$
\begin{IEEEeqnarray}{rCl}
j
\begin{pmatrix}
\frac{\partial }{\partial t} & -q\\
-q^* & -\frac{\partial }{\partial t} 
\end{pmatrix}
v
=\lambda
v.
\label{eq:Lv=lambda.v}
\end{IEEEeqnarray}
In the time domain, one can replace the
time derivative $\partial/\partial t$ by the central finite-difference matrix 
\begin{IEEEeqnarray*}{rCl}
D=\frac{1}{2\epsilon}
\begin{pmatrix}
0 & 1 & 0& \cdots & -1 \\
-1 & 0& 1 & \cdots &0\\
 & &  & \cdots &\\
0 & 0& -1 & 0 &1\\
1 & 0& 0 & -1 &0\\
\end{pmatrix},
\end{IEEEeqnarray*}
and expand \eqref{eq:Lv=lambda.v} as
\begin{IEEEeqnarray}{rCl}
j
\begin{pmatrix}
D & -\diag(q[k])\\
-\diag(q^*[k])& -D 
\end{pmatrix}
v[k]
=\lambda
v[k].
\label{eq:fd-eigp}
\end{IEEEeqnarray}
The point spectrum is contained in the eigenvalues of the matrix in the
left hand side of \eqref{eq:fd-eigp}.

Eigenvalues of a real symmetric or Hermitian matrix $A$ can be found
relatively efficiently, owing to the existence of a complete orthonormal
basis and the stability of the eigenvalues. In this case, a unitary 
matrix $P$ can be designed such that the sequence of similarity transformations $A_{k+1}=P^TA_kP$, 
$A_0=A$, converges to an almost diagonal (or a triangular) matrix.  The similarity transformation can be found
using, \eg, the QR factorization or the Householder transformation \cite{wilkinson1988aep}.

Unfortunately, most of the useful statements about computations using Hermitian
matrices cannot usually be generalized to non-Hermitian matrices. As a
result, the eigenvalues of a non-Hermitian matrix (corresponding to a
non-self-adjoint operator) are markedly difficult to calculate 
\cite{wilkinson1988aep,wilkinson1986hac}. Running
a general-purpose eigenvalue calculation routine on \eqref{eq:fd-eigp}
may therefore not be the most efficient way to get eigenvalues
($\mathcal{O}(N^3)$). One can 
study the general-purpose numerical 
eigenvalue algorithms in \cite{wilkinson1988aep} and tailor them to the discrete Zakharov-Shabat
spectral problems in this paper. Note that matrix $D$ is (almost)
tridiagonal. Thus the eigenproblem $Ax=\lambda x$ for
\eqref{eq:fd-eigp}, or equivalently discretization
of \eqref{eq:a-ode} gives a second-order recursive equation
$z_{n+2}=\alpha(\lambda)z_{n+1}+\beta(\lambda)z_n$. One can then obtain
polynomial $a(\lambda):=z_N(\lambda)$ recursively in $\mathcal{O}(N^2)$ operations
and proceed to find its roots. Alternatively, $\det(\lambda I-A)$ can be
calculated recursively.

\subsubsection{Ablowitz-Ladik Eigenproblem}
\label{sec:AL}

We can also rewrite the Ablowitz-Ladik discretization
as a matrix eigenvalue
problem. Using the $L$ operator \eqref{eq:AL-L1}, we obtain
\begin{IEEEeqnarray}{rCl}
v_1[k+1]-Q[k]v_2[k]&=&zv_1[k],
\IEEEyesnumber\IEEEyessubnumber
\label{eq:Lvk=vk-1}
\\
-Q^*[k-1]v_1[k]+\alpha[k-1] v_2[k-1]&=&zv_2[k],
\IEEEyessubnumber
\label{eq:Lvk=vk}
\end{IEEEeqnarray}
which consequently takes the form
\begin{IEEEeqnarray*}{rCl}
\begin{pmatrix}
U_1 & -\diag(Q[k])\\
-\diag(Q^*[k-1])& U^{T}_2
\end{pmatrix}
v
=z
v,
\end{IEEEeqnarray*}
in which 
\begin{IEEEeqnarray*}{rCl}
U_1&=&
\begin{pmatrix}
0 & 1 & 0& \cdots & 0 \\
0 & 0& 1 & \cdots &0\\
 & &  & \cdots &\\
0 & 0& 0 & 0 & 1\\
1 & 0& 0 & 0 &0\\
\end{pmatrix},
\\
U_2&=&
\begin{pmatrix}
0 & \alpha[0] & 0& \cdots & 0 \\
0 & 0& \alpha[1] & \cdots &0\\
 & &  & \cdots &\\
0 & 0& 0 & 0 & \alpha[N-1]\\
\alpha[N] & 0& 0 & 0 &0\\
\end{pmatrix},
\end{IEEEeqnarray*}
and
$\diag(Q[k])=\diag\left(Q[0],\cdots,Q[N]\right)$, 
$\diag(Q^*[k-1])=\diag\left(Q^*[N],Q^*[0] \cdots,Q^*[N-1]\right)$.
Note that all shifting operations are cyclic, so that
all vector indices $k$ remain in the interval $0\leq k\leq N$.

Similar results can be obtained for the simplified operator \eqref{eq:AL-L} 
instead of \eqref{eq:AL-L1}. This corresponds to the above
discretization with $U_2:=U_1$ and $-Q^*[k-1]$ replaced with
$-Q^*[k]$. Similarly, one can rewrite the normalized Ablowitz-Ladik
iteration \eqref{eq:al-iter2}  as a matrix eigenvalue problem. This corresponds to 
the $L$ operator \eqref{eq:AL-L1} with the top left and bottom right entries replaced, respectively,
with $\sqrt{\alpha[k]}\mathcal{Z}$ and $\sqrt{\alpha[k-1]}\mathcal{Z}^{-1}$. Accordingly $v_1[k+1]$ and 
$\alpha[k-1]v_2[k-1]$ in \eqref{eq:Lvk=vk-1} and \eqref{eq:Lvk=vk} are replaced, respectively, 
with $\sqrt{\alpha[k]}v_1[k+1]$ and $\sqrt{\alpha[k-1]}v_2[k+1]$  and $U_1:=U_2$ in the given expressions.

\subsubsection{Spectral Method}

In the frequency domain, one can approximate derivatives with the help
of the Fourier transform. Let us assume that
\begin{IEEEeqnarray*}{rCl}
v(t)=\sum\limits_{k=-\frac{M}{2}}^{\frac{M}{2}}\begin{pmatrix}\alpha_k\\\beta_k\end{pmatrix}e^{\frac{j2k\pi
  t}{T}},\quad q(t)=\sum\limits_{k=-\frac{M}{2}}^{\frac{M}{2}}\gamma_k e^{\frac{j2k\pi t}{T}},
\end{IEEEeqnarray*}
where $T=T_2-T_1$ and $M+1$ is the maximum number of frequencies in
all variables. Then the Zakharov-Shabat system is
\begin{IEEEeqnarray*}{rCl}
-\alpha_k\frac{2\pi k}{T}-j\sum\limits_{m=-\frac{M}{2}}^{\frac{M}{2}}\gamma_{k-m}\beta_m&=&\lambda \alpha_k, \\
-j\sum\limits_{m=-\frac{M}{2}}^{\frac{M}{2}}\gamma_{-k+m}^*\alpha_{m}+\beta_k\frac{2\pi k}{T}&=&\lambda \beta_k.
\end{IEEEeqnarray*}
Thus we obtain
\begin{IEEEeqnarray*}{rCl}
\begin{pmatrix}
\Omega & \Gamma\\
-\Gamma^H& -\Omega
\end{pmatrix}
\begin{pmatrix}
\alpha\\
\beta
\end{pmatrix}=
\lambda
\begin{pmatrix}
\alpha\\
\beta
\end{pmatrix},
\end{IEEEeqnarray*}
where
$\alpha=[\alpha_{-\frac{M}{2}},\ldots,\alpha_{\frac{M}{2}}]^T$,
$\beta=[\beta_{-\frac{M}{2}},\ldots, \beta_{\frac{M}{2}}]^T$,
$\Omega=-\frac{2\pi}{T}\diag(-\frac{M}{2},\ldots,\frac{M}{2})$
and
$\Gamma$ is a $(M+1)\times(M+1)$ Toeplitz matrix with the first row
\begin{IEEEeqnarray*}{rCl}
-j\begin{pmatrix}
\gamma_0 & \gamma_{-1} & \cdots & \gamma_{-\frac{M}{2}}& \smash[b]{\zerovector{\frac{M}{2}}}
\end{pmatrix},
\end{IEEEeqnarray*}
and the first column
\begin{IEEEeqnarray*}{rCl}
-j\begin{pmatrix}
\gamma_0 & \gamma_{1} & \cdots & \gamma_{\frac{M}{2}}& \smash[b]{\zerovector{\frac{M}{2}}}
\end{pmatrix}^T.
\end{IEEEeqnarray*}

\vspace{0.2cm}
The point spectrum is thus found by looking at the eigenvalues of the matrix
\begin{IEEEeqnarray*}{rCl}
  A=
\begin{pmatrix}
\Omega & \Gamma\\
-\Gamma^H & -\Omega
\end{pmatrix}.
\end{IEEEeqnarray*}

\section{Running Time, Convergence and Stability of the Numerical Methods}

The
numerical methods discussed in this paper are first-order matrix iterations
and therefore the running time of all of them to get $v[N]$ is $\mathcal{O}(N)$
multiplications and additions per eigenvalue. This corresponds to a
complexity of $\mathcal{O}(N^2)$ operations for the calculation of
the continuous spectrum on a mesh with $N$ eigenvalues. The exact number of operations depends on the details of the implementation and the memory requirement of the method. All iterative methods thus take about the same time asymptotically, albeit with different coefficients. 

An important observation is that, while the Fast Fourier Transform
(FFT) takes $\mathcal{O} (N\log_2 N)$ operations to calculate
the spectral amplitudes of a vector with length $N$ at $N$ equispaced frequencies, the complexity of the methods described
in this paper to compute the continuous spectrum are $\mathcal{O} (N^2)$.
Similarly, it takes $\mathcal{O} (\mathsf{N}N)$ operations to calculate the discrete spectrum. In other words, so far we do not exploit the potentially repetitive operations in our computations.

It is evident from \eqref{eq:ab-formula-a} that as
$T_2\rightarrow\infty$, $v_1[k]$ should grow as $\sim \exp(-j\lambda T_2)$ so that
$a(\lambda)$ is a finite complex number. The
canonical eigenvector $v[k;T_1,T_2]$ thus has an unbounded component as $T_2\rightarrow\infty$ (\ie, $\norm{v[k]}\rightarrow\infty$). One can, however, normalize $v_1$ and $v_2$ according to
\begin{IEEEeqnarray*}{rCl}
u_1&=&v_1e^{j\lambda t},\\
u_2&=&v_2e^{-j\lambda t},
\end{IEEEeqnarray*}
and transform \eqref{eq:dv-dt} to
\begin{IEEEeqnarray}{rCl}
u_t=
\begin{pmatrix} 
0 & q(t)e^{2j\lambda t} \\
-q^*(t)e^{-2j\lambda t} & 0
\end{pmatrix}u,\:\:
u(T_1,\lambda)=
\begin{pmatrix}1\\0\end{pmatrix}.
\label{eq:dv-dt-n}
\IEEEeqnarraynumspace
\end{IEEEeqnarray}
The desired coefficients are simply $a(\lambda)=u_1(T_2)$ and
$b(\lambda)=u_2(T_2)$. Consequently, if one is interested in obtaining
eigenvectors $v[k]$ in addition to the coefficients $a(\lambda)$ and
$b(\lambda)$, the discretization of the normalized
system \eqref{eq:dv-dt-n} has better 
numerical properties:
\begin{IEEEeqnarray}{rCl}
\begin{pmatrix}
a[k+1]\\
 b[k+1]
\end{pmatrix}
&=&\begin{pmatrix}
1 & Q[k]z^{-2k}\\
-Q^*[k]z^{2k} & 1
\end{pmatrix}
\begin{pmatrix}
a[k]\\
b[k]
\end{pmatrix},
\label{eq:vkn}
\\
\begin{pmatrix}
a[0]\\
 b[0]
\end{pmatrix}
&=&
\begin{pmatrix}
1\\
 0
\end{pmatrix}.
\nonumber
\end{IEEEeqnarray}
The nonlinear Fourier coefficients are obtained as $a:=a[N]$ and
$b:=b[N]$. The discrete nonlinear Fourier transform mentioned in
\cite{tao2006nfa} is thus the forward discretization of the normalized
Zakharov-Shabat system \eqref{eq:dv-dt-n}.

We are interested in the convergence of $v[k]$ (or $a(\lambda)$ and
$b(\lambda)$) as a function of $N$ for fixed values of $T_1$ and
$T_2$. That is to say, we require that the error
$e=\norm{v(T_1+k\epsilon)-v[k]}\rightarrow 0$ as $N\rightarrow\infty$ (for
fixed $T_1$ and $T_2$). The (global) error in all methods described in
this paper is at least $\mathcal{O}(\epsilon)$, and therefore all these methods
are convergent.

Some of these methods are, however, not stable. This is partially
because the Zakharov-Shabat system can have unbounded solutions for
$\lambda\in\Complex^+$, \ie, $\norm{v(t)}\rightarrow\infty$ as
$t\rightarrow\infty$. Errors can potentially be amplified by the
system unstable dynamics. One should be cautious about the
normalized system \eqref{eq:dv-dt-n} as well. For example,
forward discretization of the
normalized system \eqref{eq:vkn} gives a first-order iteration $x[k+1]=A_{fn}[k]x[k]$.  The
eigenvalues
of the matrix $A_{fn}[k]$ in this method are
\begin{IEEEeqnarray*}{rCl}
s_{1,2}=1\pm j\epsilon|q[k]|.
\end{IEEEeqnarray*}
It follows that the forward discretization of \eqref{eq:dv-dt-n} gives rise to eigenvalues
outside of the unit disk, 
$|s|>1$. As a result, first-order discretization of \eqref{eq:dv-dt-n}
are also unstable. In cases where $|s|>1$, we can consider normalizing the
iterations by dividing $A_{fn}[k]$ by $\sqrt{\det A_{fn}[k]}$ (in the case of
\eqref{eq:vkn}, dividing the right-hand side by
$\sqrt{1+|Q[k]|^2}$). The resulting iteration has eigenvalues inside the unit disk. For $\epsilon\ll 1$ the effect is only
second order in $\epsilon$, however it helps in managing the numerical
error if larger values of $\epsilon$ are chosen.

An issue pertinent to numerical methods is chaos. Chaos and numerical instability of finite-difference discretizations has been
observed in \cite{bishop1988qpr} for the sine-Gordon equation, which is
also integrable and shares a number of basic properties with the NLS
equation. In \cite{ablowitz1990hsn}, the authors conclude that the
standard discretizations of the cubic nonlinear Schr\"odinger equation
may lead to spurious numerical behavior. This instability is related to the homoclinic orbits of the NLS equation, \ie, it occurs
if the initial signal $q(t,0)$ is chosen to be close to the homoclinic orbit of the equation. It disappears only
if the step size is made sufficiently small, which can be smaller than
what is desired in practice.

It is shown in \cite{ablowitz1990hsn} that the Ablowitz-Ladik
discretization of the NLS equation (in time) has the desirable property that
chaos and numerical instability, which are sometimes
present in finite-difference discretizations of the NLS equation, do not
appear at all. Though these results are for the original
time-domain equations, the issue can occur in the spectral
eigenvalue problem \eqref{eq:dv-dt} as well, if the signal $q[k]$ is close to a
certain family of functions (related to $\sin\omega t$ and $\cos\omega
t$). Therefore, among discretizations studied, the Ablowitz-Ladik
discretization of the Zakharov-Shabat system is immune to chaos and 
numerical instability.  This is particularly important in the
presence of amplifier noise, where chaos can be more problematic.

\section{Testing and Comparing the Numerical Methods}
\label{sec:test-comparison}

In this section, we test and compare the ability of the suggested
numerical schemes to estimate the nonlinear Fourier transform (with
respect to the Zakharov-Shabat system) of various signals.  Numerical
results are compared against analytical formulae, in a few cases where
such expressions exist.  Our aim is to compare the speed and the precision
of these schemes for various pulse shapes in order to determine which
ones are best suited for subsequent simulation studies.

To derive the analytical formulae,
recall that the continuous spectral function can be
written as
$\hat{q}(\lambda)=\lim\limits_{t\rightarrow\infty}y(t,\lambda)$,
in which $y(t,\lambda)$ satisfies 
\cite{yousefi2012nft1}
\begin{IEEEeqnarray}{rCl}
\begin{cases}
\dfrac{\der y(t,\lambda)}{\der t}+q(t)e^{2j\lambda t}y^2(t,\lambda)+q^*(t)e^{-2j\lambda t}=0,\\
y(-\infty,\lambda)=0.
\end{cases}
\label{eq:qhat-ode}
\end{IEEEeqnarray}
Similarly, one can solve the second-order differential equation 
\begin{IEEEeqnarray}{rCl}
\begin{cases}
\IEEEstrut[1mm][0.45cm]
\dfrac{\der^2z(t,\lambda)}{\der t^2}-\left(2j\lambda+\dfrac{q_t}{q}\right)\dfrac{\der z(t,\lambda)}{\der t}+|q|^2z(t,\lambda) =0,\\
z(-\infty,\lambda)=1,\quad \dfrac{\der z(-\infty,\lambda)}{\der t}=0,
\end{cases}
\label{eq:a-ode}
\end{IEEEeqnarray}
and obtain $a(\lambda)=\lim\limits_{t\rightarrow\infty}z(t,\lambda)$. 
The zeros of $a(\lambda)$ form the discrete spectrum.

In the following,
the discrete spectrum is found and compared using
the following matrix-based schemes:
\begin{enumerate}
\item central-difference method;
\item spectral method;
\item Ablowitz-Ladik discretization with no normalization (AL1);
\item Ablowitz-Ladik discretization with normalization (AL2).
\end{enumerate}
In the matrix-based schemes, the entire point spectrum is
found at once by solving a matrix eigenvalue problem. 

The complete spectrum is found using search-based methods:
\begin{enumerate}
\item forward discretization method;
\item fourth-order Runge-Kutta scheme;
\item layer-peeling methods;
\item Crank-Nicolson method;
\item AL discretization;
\item AL discretization with normalization.
\end{enumerate}
In search-based methods, 
the Newton method is used together with the trace formula
to find both discrete and continuous spectra. 

Each of the following signals is sampled uniformly using
a total of $n$ samples in a time window containing $99\%$ of the signal energy.

\begin{figure}[t]
\centering
\includegraphics[scale=0.75]{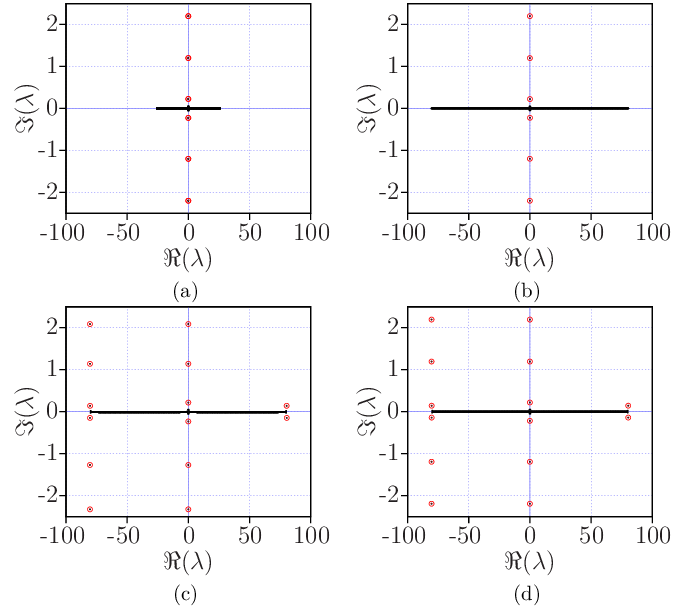}
 \caption{Discrete spectrum of the Satsuma-Yajima signal with $A=2.7$ using
(a) central-difference method,
(b) spectral method,
(c) Ablowitz-Ladik scheme,
(d) modified Ablowitz-Ladik scheme.}
\label{fig:sy-disc}
 \end{figure}

\subsection{Satsuma-Yajima Signals}

One signal with known spectrum is the
Satsuma-Yajima function  \cite{satsuma1974ivp}
\begin{equation*}
q(t)=A\sech(t),\quad A\geq 0.
\end{equation*}
Solving the initial value problem \eqref{eq:dv-dt} (or \eqref{eq:qhat-ode}) analytically, the following continuous spectral function is obtained \cite{satsuma1974ivp}
\begin{IEEEeqnarray*}{rCl}
\hat{q}(\lambda)=-\frac{\Gamma(-j\lambda+\frac{1}{2}+A)\Gamma(-j\lambda+\frac{1}{2}-A)}{\Gamma^2(-j\lambda+\frac{1}{2})}
\frac{\sin(\pi A)}{\cosh(\pi\lambda)}.
\end{IEEEeqnarray*}
The discrete spectrum is the set of zeros of $a(\lambda)$, \ie, poles of $\hat{q}(\lambda)$ (when analytically extended in 
$\Complex^+$). Recalling that $\Gamma (x)$ has no zeros and is
unbounded for $x=0,-1,-2,\ldots$, it follows that the discrete
spectrum consists of $\mathsf{N}=\lfloor A+\frac{1}{2}-\epsilon
\rfloor$ eigenvalues
\[
\lambda=\left\{(A-\frac{1}{2})j, (A-\frac{3}{2})j, \cdots \right\}, \quad \Im(\lambda)>0.
\]
In the special case in which $A$ is an integer, $A=\mathsf{N}$, the Satsuma-Yajima signal is a pure $\mathsf{N}$-soliton with $\mathsf{N}$ eigenvalues, and the continuous spectral function is zero.

\begin{figure}[t]
\centering
\includegraphics[scale=0.75]{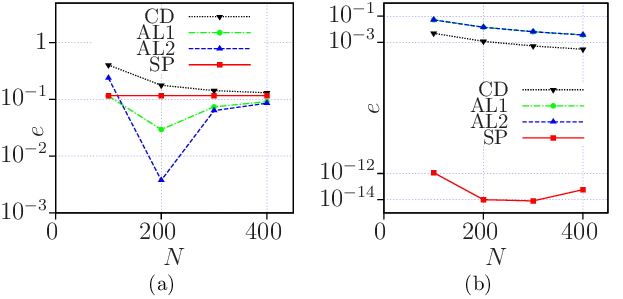} 
\caption{Error in estimating (a) the smallest eigenvalue
and (b) the largest eigenvalue of the
Satsuma-Yajima signal $q(t)=2.7\sech(t)$ as a function of the number of sample points $N$ using
matrix eigenvalue methods. 
The Ablowitz-Ladik method 1 is the method of Section \ref{sec:AL} with no
normalization, and the Ablowitz-Ladik method 2 is the same scheme with normalization.}
\label{fig:sy-error-mat}
 \end{figure}

\begin{figure}[tbp]
\centerline{
\includegraphics[scale=0.75]{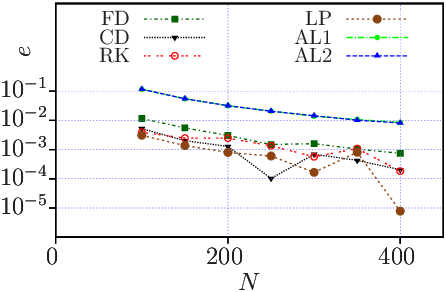}
}
\caption{Error in estimating the largest eigenvalue of Satsuma-Yajima signal $q(t)=2.7\sech(t)$ as a function of the number of sample points $N$ using search-based methods.}
\label{fig:sy-error-search}
 \end{figure}

Figs.~\ref{fig:sy-disc}, \ref{fig:sy-error-mat} and
\ref{fig:sy-error-search} give the numerical results for
$A=2.7$,  $N=2^{10}$. The value of $A$ is chosen so that both the
discrete and continuous spectrum are present. 

Fig.~\ref{fig:sy-disc} shows that, in this example, the
spectral and central-difference methods produce good results
among the matrix-based methods in estimating the discrete eigenvalues
$\lambda=0.2j, 1.2j, 2.2j$. All methods generate a large number of spurious
eigenvalues along the real axis. This behavior might be viewed as a
tendency of the algorithms to generate the continuous spectrum too.
However the spurious eigenvalues do not disappear completely even
when the continuous spectrum is absent (when $A$ is an integer);
only their range becomes more limited.
The spurious eigenvalues across the real axis can easily be filtered, since their imaginary part
has negligible amplitude. The AL methods, with and without
normalization,
produce the same eigenvalues plus another vertical line of spurious
eigenvalues having a large negative real part. Normalization in the
AL scheme does not make a significant difference in this
example.

Fig.~\ref{fig:sy-error-mat} shows 
the accuracy of the various matrix-based methods in estimating the
smallest and largest eigenvalues of $q=2.7\sech(t)$ in terms of the
number of the sample points $N$. As the number of sample points $N$ is decreased,
the spectral and central-difference methods maintain 
reasonable precisions, while the accuracy of the AL
schemes quickly deteriorates. One can check that in these cases, the 
error in the approximation
$e^{j\lambda \epsilon}\approx 1+j\lambda \epsilon$ becomes large
(since $\epsilon\gg 1$). 

It can be seen that the relative performance of the numerical methods
depends on the eigenvalue and the number of signal samples $N$.
The spectral method is generally more accurate
than the other matrix-based methods.
The AL discretizations seem to perform well as long as
$\lambda\epsilon\ll 1$, \ie, when estimating eigenvalues with small
size or when $N\geq 200$. The AL discretization eventually breaks
down at about $N=50$ as the analogy between the continuous and discrete NLS
equation is no longer justified at such low resolutions, whereas other schemes
continue to track the eigenvalues to some accuracy. In other words,
what the AL methods find at such small values of $N$ is the
spectrum of the discrete soliton-bearing 
NLS equation, which is not a feature of finite-difference
discretizations. (In fact, it is
essential for this algorithm to deviate from the finite-difference
discretizations as $N$ is reduced, to produce appropriate solitons with
few samples.) The running time of
all matrix eigenvalue methods is about the same. 

Search-based methods can be used to estimate the point spectrum as well.
Here we use the Newton method with random initial points
to locate eigenvalues in $\Complex^+$. Naturally, we limit
ourselves to a rectangular region in the complex plan, slightly above
the real axis to avoid potential spurious eigenvalues. Since the
number of eigenvalues is not known a priori, the
continuous spectrum is found first so as to give an estimate of
the energy of the discrete spectrum.  It is essential
that the continuous spectrum is estimated accurately so that
a good estimate of the energy of the discrete spectrum can
be obtained.
Once this energy is known, and
a suitable (rectangular) search region in $\Complex^+$ is chosen,
the Newton method is often able to locate
all of the discrete eigenvalues using just a few iterations. 

Fig.~\ref{fig:sy-error-search} shows the accuracy of the
searched-based methods in estimating the largest eigenvalue of the signal
$q(t)=2.7\sech(t)$. The Runge-Kutta, layer-peeling and
Crank-Nicolson methods have about the same accuracy, followed closely
by forward discretization. Since this is
the largest eigenvalue, the 
AL schemes are not quite as accurate. As noted
above, comparison at
smaller values of $N$ is not
illustrative, as the AL estimate quickly deviates from
$\lambda_{\max}$ of the continuous signal.   

The Runge-Kutta method, at the accuracies shown in the above graphs, is
of course very slow, and is not a practical method to implement. The
running time for the other schemes is approximately the same.
Search-based methods take an order
of magnitude more time than matrix-based methods when $N$ is small.
These methods fail when $N$ becomes too small
($N<200$), since the large error in estimating the energy terms of the continuous
spectrum negatively influences the stopping criteria and
consequently degrades the Newton
increments. For large $N$, on the other hand, the QR
factorization, which takes $\mathcal{O}(N^3)$ operations in
calculating the eigenvalues of a matrix, becomes quite slow and restricts the use of matrix-based
methods.

The same conclusions are observed for various choices of real or complex
parameter $A$. As $|A|$ is increased, as before, the spectral and
finite-difference schemes produce the correct eigenvalues, and the
AL methods
generate the same eigenvalues plus an additional vertical strip of spurious
eigenvalues. The range of the spurious eigenvalues across the real axis remains
about the same. As the phase of $A$ is increased, the true (non-spurious)
eigenvalues remain the same
in all methods (as expected analytically), while some of the vertical spurious
eigenvalues in the AL schemes move from left to right or
vice versa. The spectral and finite-difference
schemes are relatively immune to these additional spurious eigenvalues.
Normalization of the AL
method sometimes produces slightly fewer spurious eigenvalues across the real
axis, as can be seen in Figs.~\ref{fig:sy-disc}(c)--(d).   

\begin{figure}[tbp]
\centering
\includegraphics[scale=0.75]{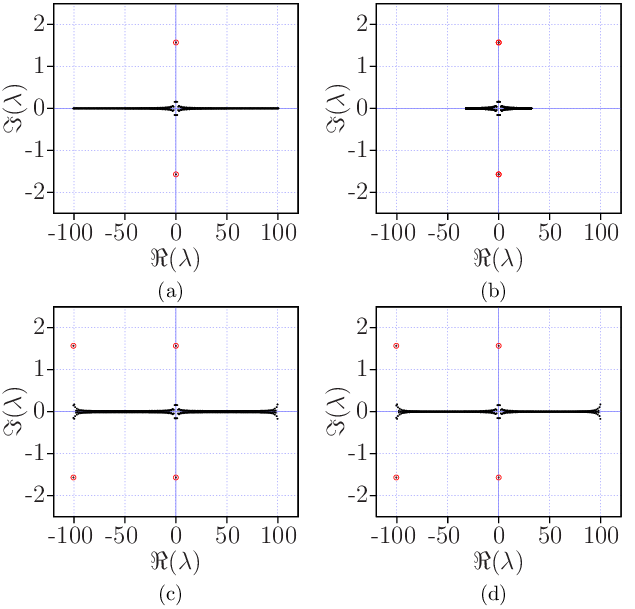}  
\caption{Discrete spectrum of the rectangular pulse
    \eqref{eq:square} with $A=2$, $T_2=-T_1=1$ using (a) spectral method,
(b) central-difference method,
(c) Ablowitz-Ladik scheme, (d) modified Ablowitz-Ladik scheme.}
\label{fig:sw-disc}
 \end{figure}

\subsection{Rectangular Pulse}

Consider the rectangular pulse
\begin{IEEEeqnarray}{rCl}
q(t)=
\begin{cases}
A, & t\in[T_1,T_2],\\
0, & \textnormal{otherwise.}
\end{cases}
\label{eq:square}
\end{IEEEeqnarray}
It can be shown that the continuous spectrum is given by \cite{yousefi2012nft1}
\begin{IEEEeqnarray*}{rCl}
  \hat{q}(\lambda)=\frac{A^*}{j\lambda}e^{-2j\lambda
    T_2}\left(1-\frac{D}{j\lambda}\cot(D (T_2-T_1))\right)^{-1},
\end{IEEEeqnarray*}
where $D=\sqrt{\lambda^2+|A|^2}$. To calculate the discrete
spectrum, the equation \eqref{eq:a-ode} is reduced to a simple constant
coefficient second-order ordinary differential equation 
\begin{IEEEeqnarray*}{rCl}
\frac{\der^2 z}{\der t^2}-2j\lambda \frac{\der z}{\der t}+|A|^2z=0,\quad z(T_1)=1, z^\prime(T_1)=0.
\end{IEEEeqnarray*}
It is easy to verify that the eigenvalues are the solutions of
\begin{IEEEeqnarray}{rCl}
e^{2j(T_2-T_1)\sqrt{\lambda^2+|A|^2}}=\frac{\lambda+\sqrt{\lambda^2+|A|^2}}{\lambda-\sqrt{\lambda^2+|A|^2}}.
\label{eq:sw-roots}
\end{IEEEeqnarray}
Following the causality and the layer-peeling property of the NFT, one
can generalize the above result to piece-wise constant pulses. This is
the basis of the layer-peeling method of Section~\ref{sec:layer-peeling}.

Figs.~\ref{fig:sw-disc}(a)--(d) show the results of numerically computing
the discrete spectrum of a rectangular pulse with parameters 
$A=2$, $T_2=-T_1=1$, denoted hereafter by $q(t)=2\rect(t)$. The exact eigenvalue is found to be
$\lambda=1.5713j$, by numerically finding the roots of \eqref{eq:sw-roots}
using the Newton-Raphson method. No other eigenvalue is found under a
large number of random initial conditions. All methods generate the desired eigenvalue together with a large 
number of spurious eigenvalues across the real axis. The central-difference
scheme visibly generates fewer spurious eigenvalues. The Ablowitz-Ladik schemes produce two more eigenvalues with a large negative real part.

\begin{figure}[tbp]
\centering
\includegraphics[scale=0.75]{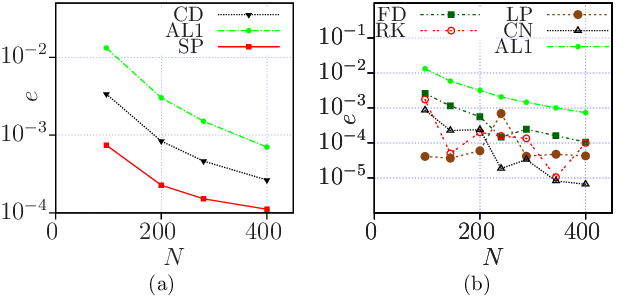} 
\caption{Error in estimating the largest eigenvalue of the rectangular 
pulse $q(t)=2\rect(t)$ as a function of the number of sample points
$N$ using (a) matrix-based methods and (b) search-based methods.}
\label{fig:sw-prec}
\end{figure}

Fig.~\ref{fig:sw-prec} compares the precision of various methods in
estimating the nonlinear spectrum of $q(t)=2\rect(t)$.
The modified AL scheme performed the same as the basic AL
scheme, and hence we  do not include the modified AL scheme in
the graphs.
\begin{figure}[tbp]
\centering
\includegraphics[scale=0.75]{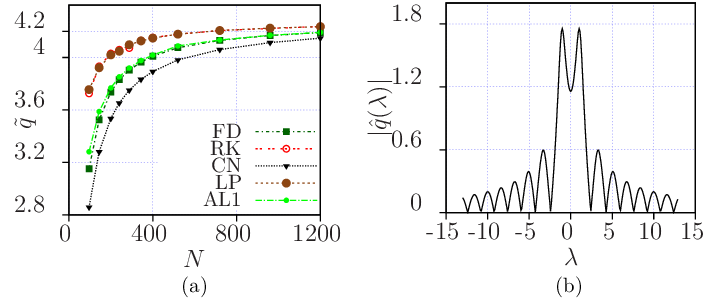}
\caption{ (a) Convergence of the discrete spectral amplitude for the rectangular pulse $q(t)=2\rect(t)$ as a function of the number of sample points $N$. Factor $-j$ is not shown in the figure. (b) Continuous spectrum.}
\label{fig:sw-C-cont-spec}
\end{figure}
\begin{figure}[tbp]
\centering
\includegraphics[scale=0.75]{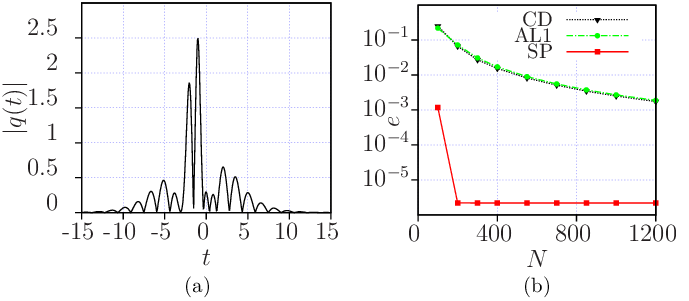}
\caption{(a) Amplitude profile of the 4-soliton signal with spectrum 
\eqref{eq:4sol}. (b) Error in estimating the eigenvalue $\lambda=1+0.5j$.}
\label{fig:4sol-pulsee}
\end{figure}

Convergence to the discrete spectral amplitudes, if it occurs at all, is generally
very slow compared with the convergence of eigenvalues. Fig.~\ref{fig:sw-C-cont-spec}(a) shows the precision of various
methods in estimating the discrete spectral amplitude of
$q(t)=2\rect(t)$. It can be seen that convergence does
not occur until $N>1000$. Fig.~\ref{fig:sw-C-cont-spec}(b)
shows the continuous spectrum
for the same function. All methods produced essentially
the same continuous spectrum,
except for some very slight variations near zero frequency.

As $|A|$ is increased, more eigenvalues appear on the imaginary axis. The
distance between these eigenvalues becomes smaller as $|\lambda|$ is
increased. All methods produce similar results, with the Ablowitz-Ladik methods
reproducing the purely imaginary eigenvalues at spurious locations
with large real part.
Phase addition has no influence on any of these methods, as
expected analytically. 
 
\subsection{$\mathsf N$-Soliton Signals}

\begin{figure}[tbp]
\centering
\includegraphics[scale=0.75]{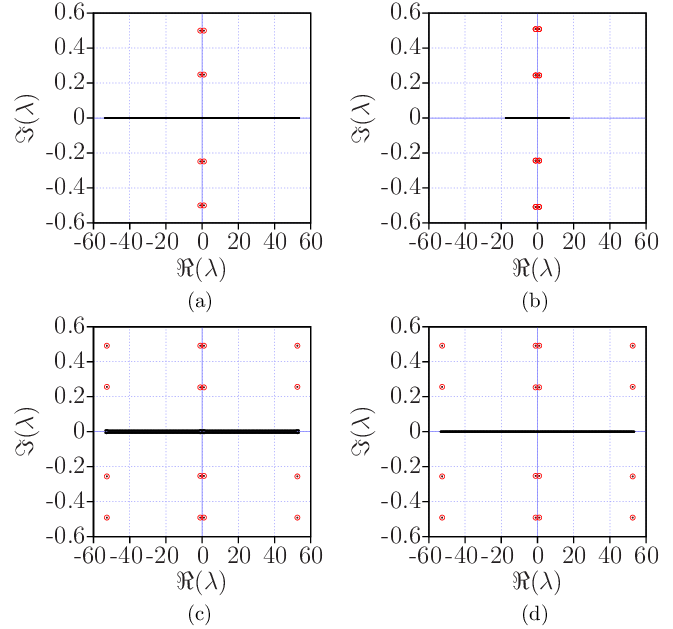} 
\caption{Discrete spectrum of the 4-soliton signal with
spectrum \eqref{eq:4sol} using (a) Fourier method,
(b) central-difference method, (c) Ablowitz-Ladik scheme, (d) modified Ablowitz-Ladik scheme.}
\label{fig:4sol-disc-spec}
 \end{figure}

\begin{figure}[tbp]
\centering
\includegraphics[scale=0.75]{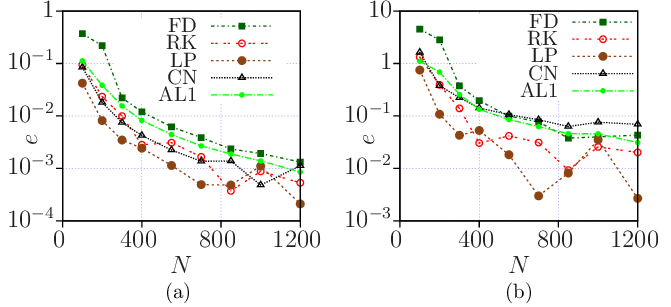}
  \caption{(a) Error in estimating the eigenvalue $\lambda=-1+0.25j$ in
    a 4-soliton using search-based methods. (b)
Error in estimating the discrete spectral amplitude $|\tilde{q}|=1$.}
\label{fig:4sol-search}
 \end{figure}

We consider a $4$-soliton signal with discrete spectrum 
\begin{IEEEeqnarray}{rClrCl}
\tilde{q}(-1+0.25j)&=&1, \quad&\tilde{q}(1+0.25j)&=&-j, \nonumber\\
\tilde{q}(-1+0.5j)&=&-1,\quad&\tilde{q}(1+0.5j)&=&j.
\label{eq:4sol}
\end{IEEEeqnarray}
The $4$-soliton is generated by solving the
Riemann-Hilbert linear system of equations with
zero continuous spectrum \cite{yousefi2012nft1} and can be seen
in Fig.~\ref{fig:4sol-pulsee}(a). 
Figs.~\ref{fig:4sol-disc-spec}(a)--(d) show the discrete spectrum
of the signal using various matrix-based methods. The relative accuracy
of these schemes in estimating the eigenvalue $\lambda=1+0.5j$
is shown in Fig.~\ref{fig:4sol-pulsee}(b). A very similar graph
is obtained for other eigenvalues. Iterative methods fare similarly
and their performance is shown in Figs.~\ref{fig:4sol-search}(a)--(b). 


The convergence of the discrete spectral amplitudes
$\tilde{q}(\lambda_j)$ is not quite satisfactory (Fig.~\ref{fig:4sol-search}(b)).
Discrete spectral amplitudes associated with eigenvalues with small
$|\Im(\lambda_j)|$ can be obtained with reasonable accuracy, although
the convergence of $\tilde{q}(\lambda_j)$ is slower than the convergence of the eigenvalues themselves.
On the other hand, discrete
spectral amplitudes associated with eigenvalues with large $|\Im(\lambda)|$ are
extremely sensitive to the location of eigenvalues and even slight changes
in eigenvalues lead to radically different estimates for the spectral
amplitudes. In fact, as the energy of the signal is increased by having
eigenvalues with large $|\Im(\lambda)|$, the Riemann-Hilbert system becomes
ill-conditioned. Therefore the discrete spectral amplitudes cannot generally be
obtained using the methods discussed in this paper. It is illustrative to see
the surface of $|a(\lambda)|$ in Fig.~\ref{fig:pulse-prop}. The eigenvalues
sometimes correspond to deep and narrow wells in the surface of $|a(\lambda)|$,
and sometimes they correspond to flat minima. In cases that they correspond to
narrow wells, the derivative $a^\prime(\lambda)$ is sensitive to the location
of eigenvalues, leading to sensitivities in $\tilde{q}(\lambda_j)$.

\begin{figure}[tbp]
\centering
\includegraphics[scale=0.75]{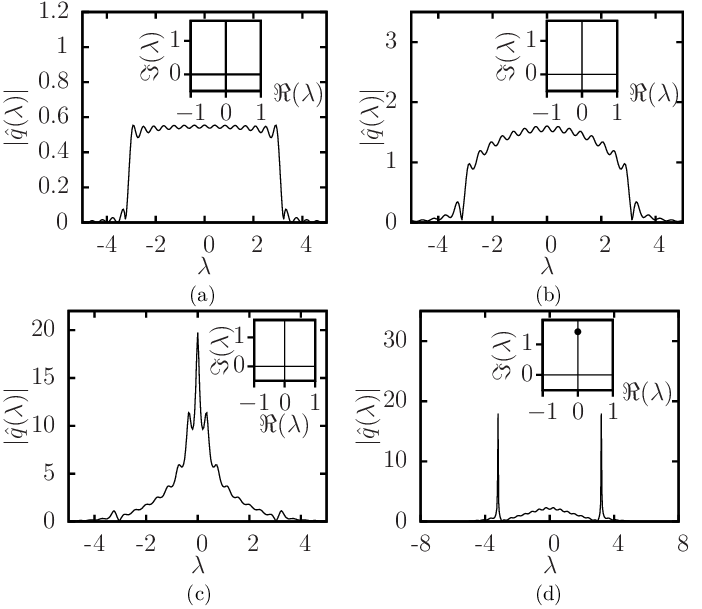}
\caption{Nonlinear Fourier transform of a 
sinc function with
amplitude a) $A=1$, b) $A=2$, c) $A=3$, d) $A=4$.}
\label{fig:sinc-a1}
 \end{figure}

Note that
$\tilde{q}(\lambda_j)$ do not appear in the trace formula, and in particular
they do not contribute to the signal energy.  This part of the NFT controls the
time center of the signal and influences the signal phase and shape too. Due to dependency
on the time center of the signal and the fact that time center can hardly be
used for digital transmission, the values of $|\tilde{q}(\lambda_j)|$ in
the Riemann-Hilbert approach appear to
be numerically chaotic and cannot carry much information. For this reason, we
do not discuss these quantities in detail.

\section{Nonlinear Fourier Transform of Signals in Data Communications}

In this section, we use the numerical methods discussed in Section~\ref{sec:numerical-methods} to compute
the nonlinear Fourier transform
of signals typically used in optical fiber transmission. The emphasis is on sinc functions
as they constitute signal degrees-of-freedom, but we also consider
raised-cosine, $\sech(\omega t)$ and Gaussian signals. In particular, we study the effect of the amplitude and phase
modulation on the structure of the nonlinear spectra. We will also discuss the spectrum of wavetrains formed
by sinc functions. 

Since the layer-peeling and the spectral methods give accurate results in estimating the nonlinear
spectra, they are chosen for subsequent simulations. 

\begin{figure}[tbp]
\centering
\includegraphics[scale=0.75]{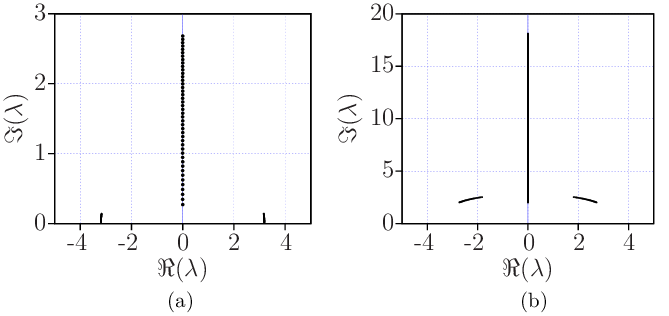}
\caption{Locus of eigenvalues of the sinc function under amplitude 
modulation: (a) $A=0$ to $A=5$, (b) $A=0$ to $A=20$.}
\label{fig:sinc-locus}
 \end{figure}
\begin{figure}[tbp]
\centering
\includegraphics[scale=0.75]{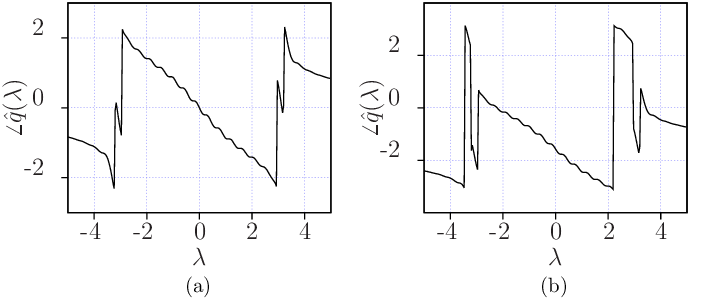} 
  \caption{Phase of the continuous spectrum of a sinc function when:
(a) $A=4$ (b) $A=4j$.}
\label{fig:sinc-phase}
 \end{figure}

\begin{figure*}[t!]
\centering
\includegraphics[scale=0.75]{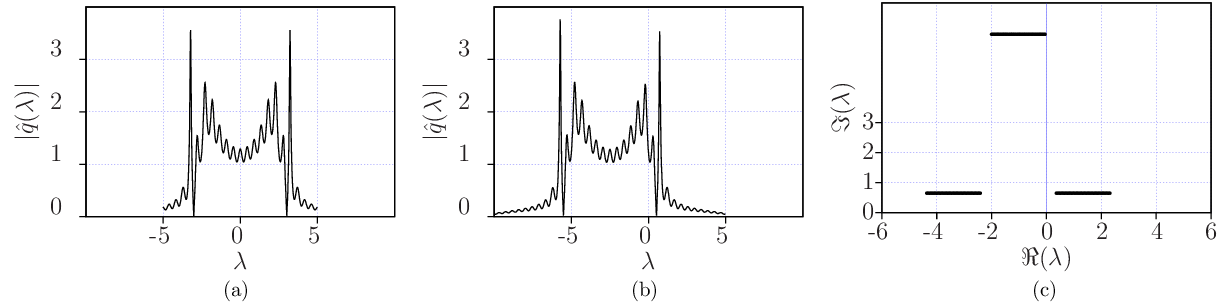}
  \caption{ (a) Amplitude of the continuous spectrum with no
    carrier. (b) Amplitude of the continuous spectrum with carrier frequency
    $\omega=5$. The phase graph is also shifted similarly with no
    other change ($\Delta\lambda=2.5$). (c) Locus of the eigenvalues of a sinc function with amplitude $A=8$ as the carrier frequency 
 $\omega$ in $\exp(-j\omega t)$ varies.}
\label{fig:sinc-lin-chirp}
 \end{figure*}

\subsection{Spectra of the Nyquist Functions}
\subsubsection{Amplitude and Phase Modulation of Sinc Functions}

Fig.~\ref{fig:sinc-a1} shows the spectrum of $q(t)=A\sinc(2t)$ under the
amplitude modulation.  Note that this function does not satisfy the
Assumption~\ref{ass:A}(a) since $q(t)\notin L^1(\Reals)$. We therefore
assume that the signal is truncated in a sufficiently large finite time
interval. It can be seen that the nonlinear Fourier
transform of $q(t)$ is all dispersive as
$A$ is increased from zero, until about $A=\pi$ where a new
eigenvalue emerges from the origin. Starting from $A=0$, the continuous spectrum is a rectangle, resembling the ordinary
Fourier transform $-\mathcal{F}(q^*(t))(2\lambda)$. As $A$ is increased, the continuous spectral function is narrowed
until $A=\pi$, where it looks like a delta function and its energy
starts to deviate from the energy of the time-domain signal.
As $A>\pi$ is further increased, the dominant eigenvalue on the
$j\omega$ axis moves up until $A=1.27\pi$, where $\lambda_1=1.4234j$
and a new pair of eigenvalues emerges, starting from $\lambda_{23}=\pm
3.2+0.05j$. When the newly created eigenvalues are not pronounced
enough, for instance in this example when transiting from $A=\pi$ to
$A=1.27\pi$, numerical algorithms have difficulties in determining
whether these small emerging eigenvalues are part of the spectrum or
not. Here it appears that for $\pi<A\leq 1.27\pi$ there is just one
dominant purely imaginary eigenvalue moving upward. At $A=1.27\pi$,
$\lambda_{23}$ emerge and move up in the complex plane as $A$ is
increased. An important observation is that the sinc function appears
to have not only purely imaginary eigenvalues, but also a pair of
symmetric eigenvalues with nonzero real part emerging at high values of $A$;
see Fig.~\ref{fig:sinc-locus}(b). This means that, for example, a
sinc function (viewed in the time domain) contains a stationary
``central component'' plus two small ``side components'' which travel
to the left and right if the sinc function is subject to the NLS
flow. The locus of the eigenvalues of the function $A\sinc(2t)$ as a result of variations
in $A$ is given in Fig.~\ref{fig:sinc-locus}.

It follows that the (truncated) sinc function is a simple example of a real symmetric signal
whose eigenvalues are not necessarily purely imaginary, as conjectured
for a long time \cite{klaus2003ezss}. 
However if $q(t)$ is real, non-negative, and ``single-lobe'', then there
are exactly $\mathsf{N}=\lfloor
\frac{1}{2}+\frac{||q||_{L_1}}{\pi}-\epsilon\rfloor$ eigenvalues, all
purely imaginary \cite{klaus2003ezss}. 

\begin{remark}
It appears that the (truncated) sinc function has a large number of
eigenvalues distributed around the real
axis. Since these eigenvalues have small imaginary parts and there are
too many of them, sometimes it is hard
to tell if they are spurious or part of the spectrum. This portion of
the spectrum can be sensitive to the number of signal samples $N$ and
the truncation window as well. One should thus make
sure that $N$ is large enough and
eigenvalues are computed and compared using both search and matrix-based
methods. Generally speaking, only dominant eigenvalues with large enough
imaginary parts can be trusted. The continuous spectrum also appears
somewhat
oscillatory. As a result, sinc functions do not appear
to have desirable nonlinear spectrum.
\end{remark}

\begin{figure}[tbhp]
\centering
\includegraphics[scale=0.75]{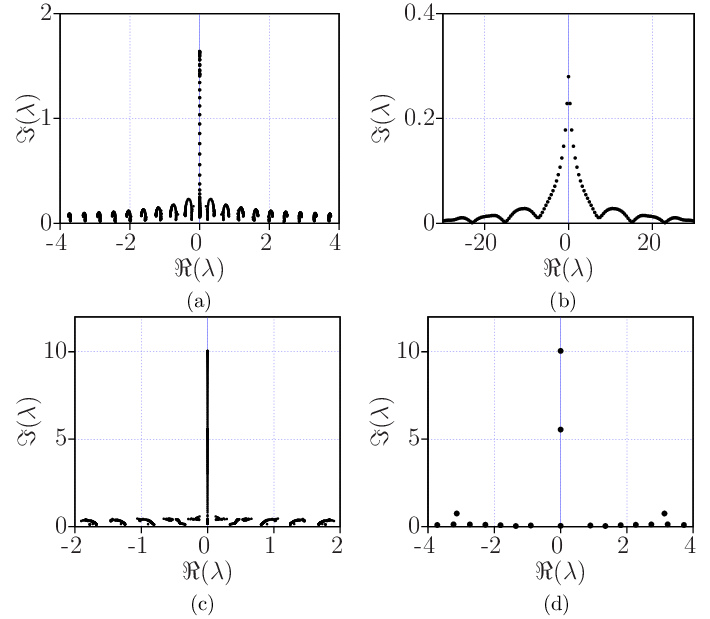}
\caption{Eigenvalues of $Ae^{j\omega t^2}\sinc(2t)$: (a) locus of eigenvalues
for $A=4$ and $0.5\leq \omega\leq 50$, (b) 
eigenvalues for $A=4$ and $\omega=15$, (c) locus of eigenvalues for $A=12$
and $0\leq \omega \leq 50$,  (d)
eigenvalues for $A=12$ and $\omega=0.50$.
}
\label{fig:sinc-quad-chirp}
\end{figure}

Under phase modulation, in the form of introducing a constant phase term
to the signal, the eigenvalues and the magnitude of the continuous spectrum
remained unchanged. Vertical (negative) shift in the phase of the continuous spectrum as a result
of phase modulation can be seen in Fig.~\ref{fig:sinc-phase}.

We may also examine the effect of time-dependent phase changes.
The effect of linear chirp, of the form $\exp(j\omega t)$, is shown
in picture Fig.~\ref{fig:sinc-lin-chirp}. Linear chirp results in
just a shift of the discrete and continuous spectrum to the left or the right,
depending on the sign of the chirp. 

It is interesting to observe the effect of a quadratic chirp. The locus of
eigenvalues that result due to changes in the quadratic phase $q\exp(j\omega t^2)$
has
been studied in \cite{klaus2003ezss} for Gaussian pulses. In our sinc function
example, in the case that there is one discrete eigenvalue in the chirp-free
case (such as when $A=4$), increasing $\omega$ will move the eigenvalue on the
$j\omega$ axis upward, but then the eigenvalue moves down again and is absorbed
in the real axis; see Figs.~\ref{fig:sinc-quad-chirp}(a)--(b). Note that the eigenvalues off the $j\omega$
axis with small imaginary parts are considered to be spurious; their number increases as the
number of sample points is increased.

A more interesting behavior is observed when $A=12$. Here, there are two
eigenvalues on the $j\omega$ axis: $\lambda_1\approx 10.05j$,
$\lambda_2=5.55j$, together with $\lambda_{3,4}=\pm 3.1 +
0.18j$ (Fig~\ref{fig:sinc-quad-chirp}(d)). As
$\omega$ is increased from zero, $\lambda_1$ and $\lambda_2$ move down and a fifth
eigenvalue $\lambda_5$ emerges from the real axis and moves upwards on the $j\omega$ axis.
It appears that at
about $\omega=41.41$, $\lambda_2$ and $\lambda_5$ ``collide'' and move out
of the $j\omega$ axis to the left and right. If $\omega$ is further increased,
$\lambda_2$ and $\lambda_5$ are absorbed into the real axis; see
Figs.~\ref{fig:sinc-quad-chirp} (c)--(d).   
\begin{figure*}[tbp]
\centering
\includegraphics[scale=0.75]{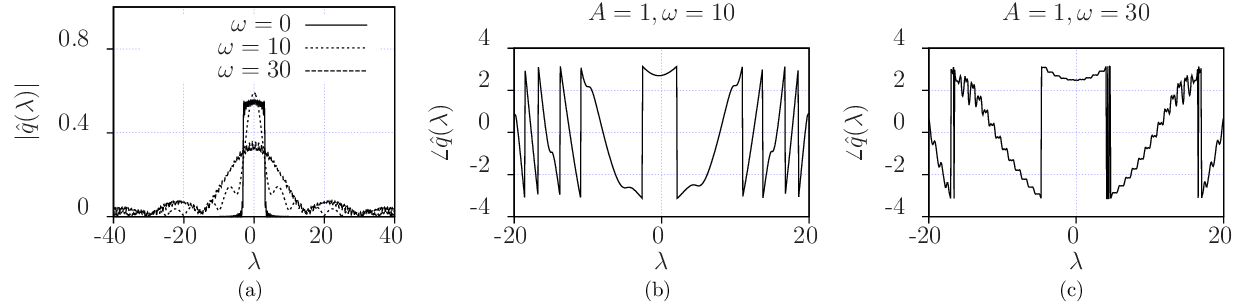}
\caption{(a)
Nonlinear spectral broadening as a result of quadratic phase modulation $Ae^{j\omega t^2}\sinc(2t)$ with $A=1$ and $\omega=0$, $10$ and $30$.
(b) Phase of the continuous spectrum when $A=1$ and $\omega=10$. (c) Phase of the continuous spectrum when $A=1$ and $\omega=30$.}
\label{fig:sinc-quad-chirp-rho}
 \end{figure*}

\begin{figure}[t]
\centering
\includegraphics[scale=0.75]{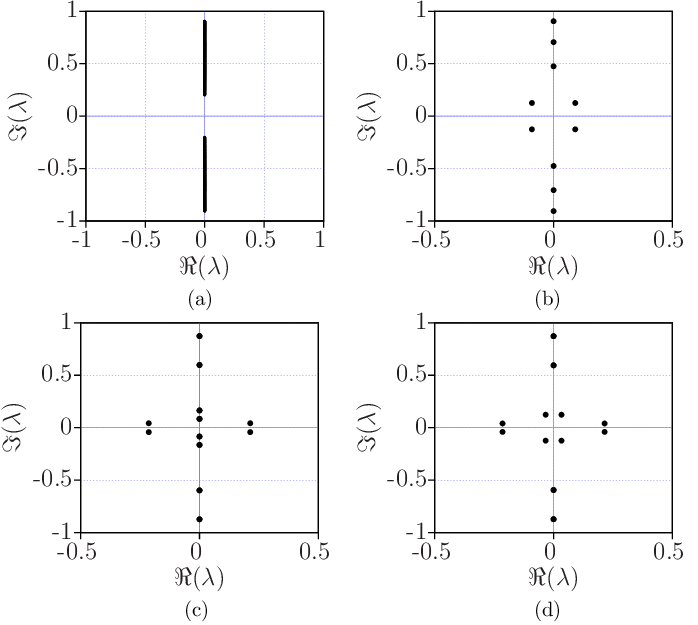} 
\caption{ Locus of eigenvalues of $\sinc(at)$ as the bandwidth $a$
varies (a) from $a=0.1$ to $a=0.6$. (Eigenvalues with small $\Im\lambda$
are not shown here.) (b) Eigenvalues for $a=0.1$. (c) Eigenvalues
before collision and (d) after collision.
}
\label{fig:sinc-dilation}
 \end{figure}

The locus of eigenvalues as a function of the bandwidth can be studied similarly. Signal
$q(t)=\sinc(at)$ has 3 eigenvalues on the $j\omega$ axis for $a=0.1$, plus two
small eigenvalues on two sides of the $j\omega$ axis
(Fig~\ref{fig:sinc-dilation}(b)). As $a$ is increased, the
smaller eigenvalue on the $j\omega$ axis comes down and a new eigenvalue is
generated at the origin, moving upward. These two eigenvalues collide at
$0.12j$ ($a=0.1330$) and are diverted to the first and second quadrant,
and eventually absorbed in the real axis at about $\Re\lambda=\pm 0.32$
($a=0.1990$), Figs.~\ref{fig:sinc-dilation}(c)--(d)). As $a$ is decreased, more eigenvalues appear on the
$j\omega$ axis and fewer on the real axis. Note that the eigenvalues are not
necessarily on the $j\omega$ axis. For example, the signal $y=\sinc(0.1370 t)$
clearly has eigenvalues $\lambda_1=0.8684j$, $\lambda_2=0.5797j$,
$\lambda_{3,4}=\pm 0.1055 + 0.1210i$.

Fig.~\ref{fig:sinc-quad-chirp-rho} shows the nonlinear spectrum
of a sinc function under a quadratic chirp modulation, given by
$Ae^{j\omega t^2}\sinc(2t)$,
is broadened as $\omega$ varies.

\begin{figure}[tbp]
\centering
\includegraphics[scale=0.75]{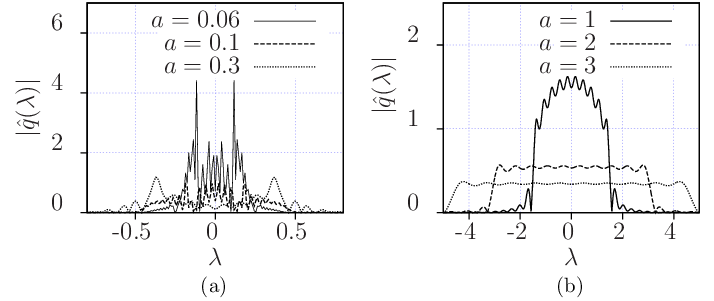}
\caption{Bandwidth expansion in $\sinc(at)$ for (a) $a=0.06$, $a=0.1$, $a=0.3$ (b) $a=1$, $a=2$, $a=3$.}
\label{fig:sinc-dilation-rho}
 \end{figure}

The effect of time dilation on the
continuous spectrum can be seen in Fig~.\ref{fig:sinc-dilation-rho}.
It can be observed that increasing bandwidth $a$, will
increase the continuous range of real nonlinear frequencies,
leading to bandwidth expansion. 

\subsubsection{Sinc Wavetrains}

The
nonlinear spectrum of a wavetrain can take on a complicated form,
just like its ordinary Fourier transform counterpart. Eigenvalues
of a two-symbol train, for instance, depend on the amplitude and phase
of the two signals, and their separation distance. 

\begin{figure}[tbp]
\centering
\includegraphics[scale=0.75]{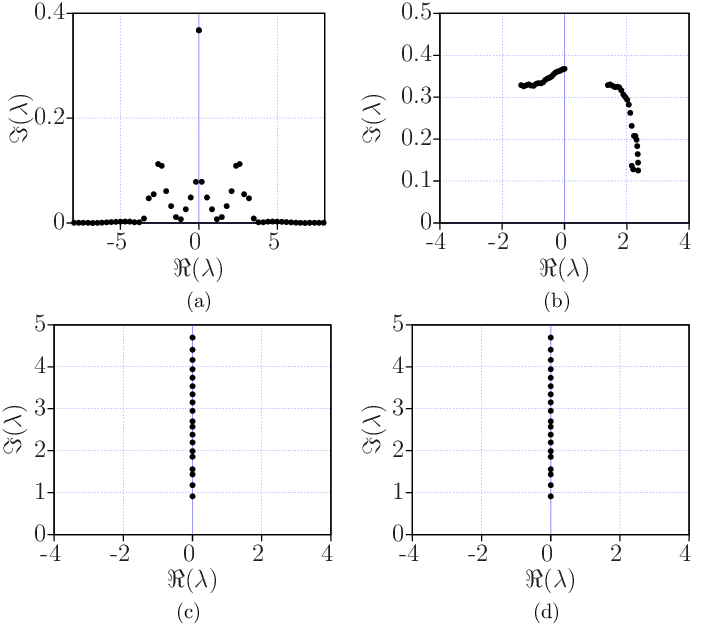} 
  \caption{Discrete spectrum of $q(t)=a_1\sinc(2t-1)+a_2\sinc(2t+1)$
    for (a) $a_1=a_2=2$, (b) $a_1=2$, $a_2=2e^{j\theta}$ for
    $-\pi<\theta \leq \pi$, (c) 
$a_1=2$, $0\leq a_2 \leq 6$, (d) $a_1=4j$ , $0 \leq a_2 \leq 6$. }
\label{fig:2sinc-1}
 \end{figure}

We first analyze the case in which there are only two sinc functions
located at the fixed Nyquist distance from each other, \ie, 
$q(t)=a_1\sinc(2t-1)+a_2\sinc(2t+1)$.
For $a_1=a_2=2$ the spectrum consists of $\lambda_1=0.3676j$,
$\lambda_{2,3}=\pm 2.5+0.03j$ and a number of spurious eigenvalues as shown in Fig.~\ref{fig:2sinc-1}(a).
As the phase of $a_2$ is increased from $\theta=0$ to $\theta=\pi$,
$\lambda_1$ moves off the $j\omega$ axis to the left and a new eigenvalue
emerges from the real axis in the first quadrant. Eigenvalues at $\theta=\pi$ are $\pm 1.3908 + 0.3287i$. The 
resulting locus of eigenvalues is shown in Fig.~\ref{fig:2sinc-1}(b).
Figs.~\ref{fig:2sinc-1}(c)--(d) depict similar graphs when $a_1$ or $a_2$ change.

\begin{figure}[t]
\centering
\includegraphics[scale=0.75]{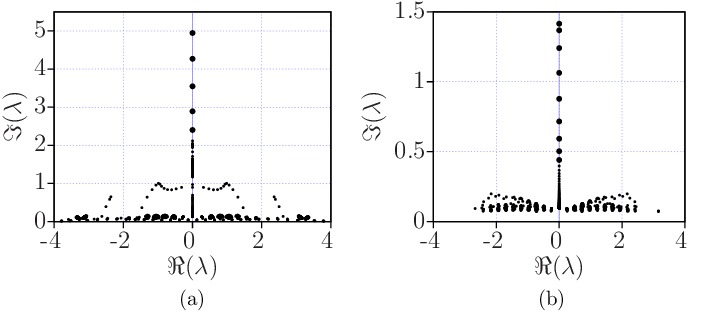} 
\caption{(a) The locus of the discrete spectrum of $q(t)=4\sinc(2t+2\tau)+4\sinc(2t-2\tau)$ as a function of $0\leq \tau\leq 5$. (b) The locus of the discrete spectrum of $q(t)=2\sinc(2t+2\tau)+2\sinc(2t-2\tau)$ as a function of $0\leq \tau\leq 5$.}
\label{fig:2sinc-delay}
 \end{figure}

Next we study the locus of the discrete spectrum as a function of pulse
separation for fixed amplitudes. If the amplitude of the sinc functions is increased sufficiently,
eigenvalues appear off the real axis and form a locus as the distance
between pulses varies.
Fig.~\ref{fig:2sinc-delay}(a) shows the locus of eigenvalues of
$q(t)=4\sinc(2t+2\tau)+4\sinc(2t-2\tau)$ as $\tau$ changes between zero to 5. At
$\tau=0$, eigenvalues are $\lambda_1=6j$, $\lambda_{2,3}=\pm 2.3618 + 0.6476i$
and $\lambda_{4,5}=\pm 3.2429 + 0.0815i$. As the distance between pulses is
increased, $\lambda_1$ rapidly decreases, and at about $\tau=0.2$,
where $\lambda_1=4.27j$, the eigenvalues with non-zero real parts are absorbed into
the real axis at $\Re\lambda=-3$. As $\tau$ is further increased, $\lambda_1$
decreases further, until $\tau=0.4$ where $\lambda_1=2.12j$ and two new
eigenvalues emerge at locations $\Re\lambda=\pm 1.5$ going up and towards the
$j\omega$ axis. There are also a large number of eigenvalues with
small imaginary parts which arise from the real axis but return, before reaching the $j\omega$
axis, to be absorbed into the real axis, while new similar eigenvalues are generated
again from the real axis. At $\tau=0.7$ eigenvalues are
$\lambda=2j,\:\pm 1+j$. At some point, the newly 
created eigenvalues are not absorbed into the
real axis, but they reach the $j\omega$ axis and collide.
For instance, at $\tau=1$ the two eigenvalues off the $j\omega$ axis rapidly travel
towards the $j\omega$ axis and collide at $\tau=1.05$.  One of these eigenvalues goes down to be absorbed
into the origin, and
the other one, interestingly, goes up to be united with the maximum eigenvalue
on the $j\omega$ axis (\ie, to create one eigenvalue with multiplicity two).
Increasing the distance further does not change the location of this
eigenvalue, which from now is fixed at $\lambda=1.4j$, but just changes the
pattern of lower level eigenvalues. At this point ($\tau=5$), the two
sinc functions are separated in time. The collision does not occur when the
amplitudes of the signals are smaller; see Fig.~\ref{fig:2sinc-delay} (b) for
the locus of the eigenvalues when the amplitude of the two sinc functions is 2. 

\begin{figure}[tbhp]
\centering
\includegraphics[scale=0.75]{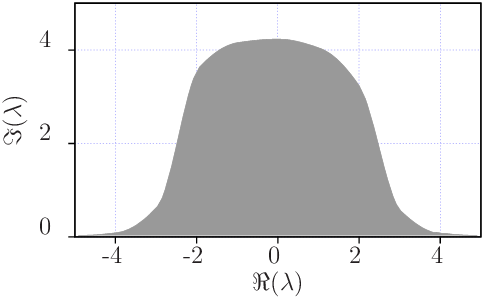} 
 \caption{Effect of the bandwidth constraint on the location of the
   eigenvalues of a sinc wavetrain containing 16 pulses with random amplitudes.}
\label{fig:sinc-trains}
 \end{figure}

\begin{remark}
When eigenvalues seem to collide or unite together, the
continuous spectrum develops delta-like spikes in its graph. This
singularity (ill-conditioned eigenproblem) makes it difficult to
determine eigenvalues accurately.   
\end{remark}

For wavetrains with a larger number of signals, the number of eigenvalues increases 
proportionally. We generate these wavetrains randomly and examine the
region to which the spectrum is confined. Fig.~\ref{fig:sinc-trains} shows the
locus of the discrete spectrum of all sinc wavetrains with 16
signals. All 16 signal 
degrees-of-freedom in the bandlimited signal are modulated here. The effect of
the bandwidth constraint in the nonlinear spectral domain can seen in
this picture.

\subsection{Preservation of the Spectrum}
It is crucial to ensure that the spectrum found by the 
numerical methods, such as those discussed in the previous sections, is in fact correct. While it proved difficult
to do so consistently and efficiently, there are various tests to
increase one's confidence in the truth of the output of the numerical
methods. Taking the inverse nonlinear Fourier transform in the
continuous-time domain and comparing the resulting function in time with the original signal is generally quite cumbersome and not always feasible. One quick test is to
examine a time frequency identity, such as the trace formula for
$n=1,2,3,\ldots$ as used in this paper. The first few conserved
quantities in this identity can be written explicitly. One should
allow higher tolerance values in the trace formula for large $n$, as the
discrete terms in this identity involve $\lambda^n$ and thus are
increasingly more sensitive to the
eigenvalues. Another test is to subject the 
signal to the flow of an integrable equation, such as the NLS
equation, and check that the discrete
spectrum is preserved and the spectral amplitudes are scaled
appropriately according to that equation. In this section, we let the signal propagate according to
the NLS equation and compare the spectra at $z=0$ and $z=\const{L}$
for various $\const{L}$. 

\begin{figure*}[p]
\centering
\includegraphics[scale=0.75]{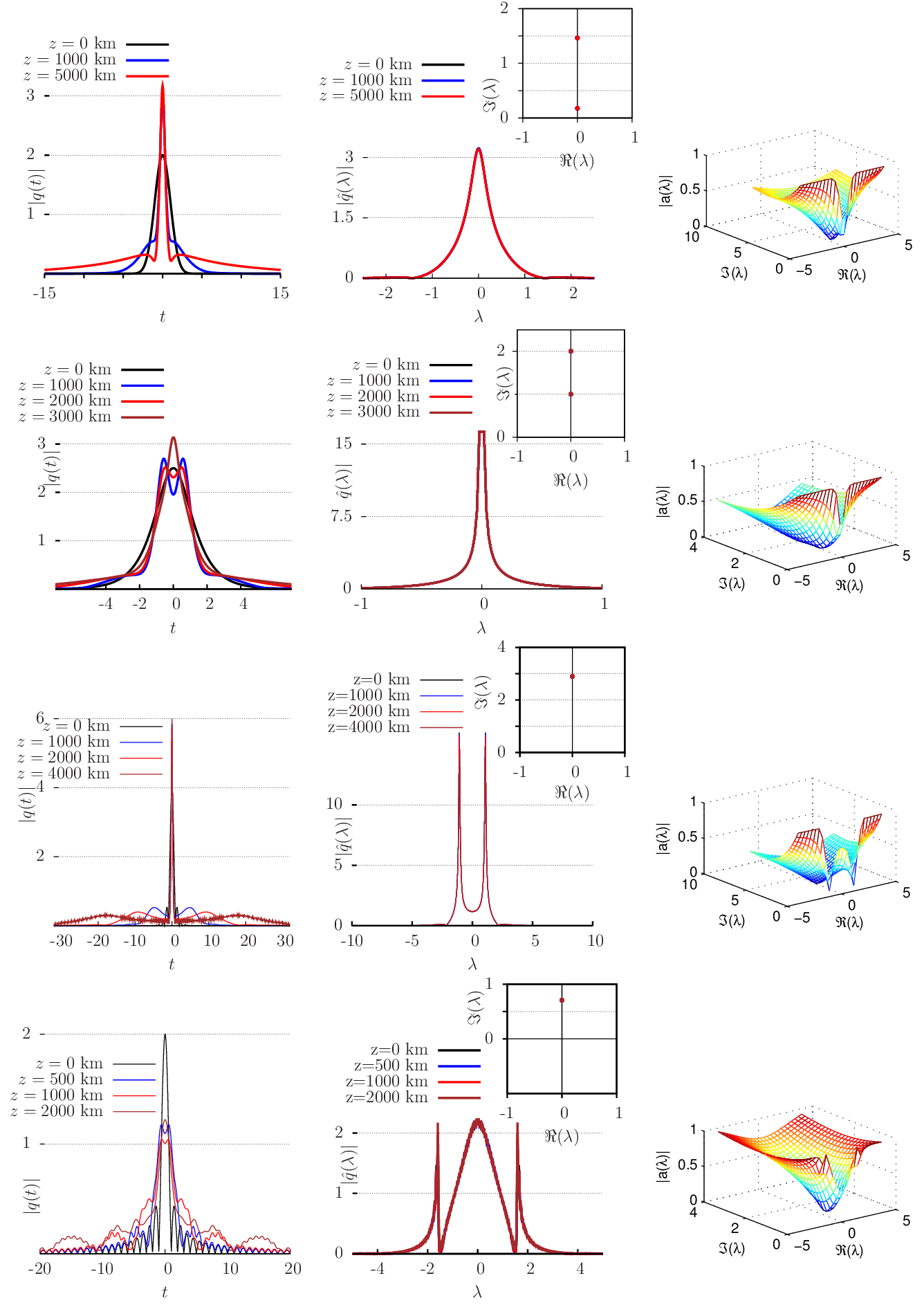}
\caption{Propagation of signals along
an optical fiber in the time domain (left),
in the nonlinear Fourier transform domain (middle), and showing the
surface of $|a(\lambda)|$ (right).  The signals are
(a) Gaussian signal,
(b) Satsuma-Yajima signal,
(c) raised-cosine function,
(d) sinc function.
The zeros of $|a(\lambda)|$ correspond
to eigenvalues in $\Complex^+$.}
\label{fig:pulse-prop}
\end{figure*}

Fig.~\ref{fig:pulse-prop} shows examples of the
spectra of a number of signals at $z=0$ and $z=\const{L}$ evolving
according to the NLS equation \eqref{eq:nlsn}. The distances
mentioned in the graphs in $\rm km$ correspond to a standard optical fiber with parameters in 
Table~\ref{tbl:fiberparam}.
\begin{table}[t]
\caption{Fiber Parameters}
\label{tbl:fiberparam}
\centerline{\begin{tabular}{c|l|l}
$D$ & $17~{\rm ps}/({\rm nm-\rm km})$ & {\footnotesize Dispersion parameter} \\
$\gamma$ & $1.27~{\rm W}^{-1}{\rm km}^{-1}$ & {\footnotesize nonlinearity parameter} \\
\end{tabular}}
\end{table}
Note that in all these examples the discrete spectrum is completely
preserved, and the continuous spectral amplitudes undergo a phase
change properly. Compared to Gaussian and raised-cosine examples, whose
nonlinear Fourier transform can be found easily, the discrete spectrum of sinc
functions is much more challenging to find. This is because the
non-dominant eigenvalues off the $j\omega$ axis have small imaginary
parts for typical parameters and are not sufficiently
distinguished. They also have large real parts, increasing the search
region. Sinc functions are thus
not the best examples to illustrate the application of the NFT in
optical fibers. We studied these ideal signals primarily because of
their fundamental utility in linear digital communications.

\section{Conclusions}

In this paper, we have suggested and compared a variety of
numerical methods for the computation of the nonlinear Fourier
transform of a signal defined on the entire real line.
A straightforward finite-difference discretization, such as the forward
discretization, does not often produce satisfactory results.
Among the methods studied in this paper, the layer-peeling
and spectral methods gave accurate results in estimating
the continuous and discrete spectrum over a wide class of
examples.

Given a waveform without having prior knowledge of the
location of the discrete eigenvalues, we suggest the use
of matrix-based methods to compute the discrete spectrum.
If, on the other hand, the location of the eigenvalues is
known approximately (as in data-communication
problems, where the eigenvalues are chosen at the
transmitter from a finite set)
a search-based method is recommended.

Although the eigenvalues and the continuous spectral function can be
calculated with great accuracy, the discrete spectral amplitudes are
quite sensitive to the location of the eigenvalues, even in the absence
of noise.  The discrete spectral amplitudes control the time center of
the signal, and are therefore sensitive to timing jitter.
For data communication purposes it follows
that, whereas the presence or absence of the eigenvalue itself may allow
for robust information transmission, encoding information
in the time center of the signal, \ie, in the discrete
spectral amplitudes (in the Riemann-Hilbert approach), is unlikely to be viable.

Using these numerical methods, we studied
the influence of various signal parameters on the
nonlinear Fourier transform of a 
number of signals commonly used in data communications.
We found, for example, that the spectrum of an isolated
normalized sinc function with amplitude $A$ is purely continuous for
the $A<\pi$. However, as the signal amplitude
is increased, dominant eigenvalues appear on the
$j\omega$ axis, together with pairs of symmetric eigenvalues having
nonzero real parts.

In general,
amplitude variations result in variations in the location
of the eigenvalues and the shape of the continuous
spectrum.  
Eigenvalues follow particular trajectories in the
complex plane.
Phase variations, on the other hand, influence only the phase
of the spectrum, not the location of the eigenvalues.
One important observation, which may be beneficial for the design of
data communication systems,
is that the nonlinear spectrum of
bandlimited signals appears to be confined to a vertical
strip in the complex plane with a width proportion to the
signal bandwidth.

This paper has only scratched the surface of a potentially
rich research area.  The development of efficient and robust
numerical techniques suitable for various engineering applications
of the nonlinear Fourier transform will require
significant additional effort.  A problem of particular
interest is the development of a ``fast'' nonlinear Fourier transform method that would be the analog of the FFT.

\end{document}